\documentclass[11pt]{article}
\usepackage{graphicx,epsfig,color}
\usepackage[english]{babel}
\usepackage{amssymb,amsfonts,amsmath,bm,bbm,bbold}
\usepackage{verbatim}

\usepackage{jheppub}
\newcommand{\bra}[1]{\langle #1|}
\newcommand{\ket}[1]{|#1\rangle}
\newcommand{\vev}[1]{\left\langle{#1}\right\rangle}

\newcommand{\dd}{{\rm d}}
\newcommand{\cO}{{\cal O}}
\newcommand{\im}{i\,}
\newcommand{\Tr}{{\rm Tr}}

\begin{document}

\title{Supergravity flows, wormholes and their pseudo-Hermitian holographic duals%
}

\author[a,b]{Jose M. Begines,}
\author[a,b]{Ant\'on F. Faedo,}
\author[a,b]{Carlos Hoyos,}
\author[a,b]{Daniele Musso}
\affiliation[a]{Department of Physics, Universidad de Oviedo, c/ Leopoldo Calvo Sotelo 18, ES-33007 Oviedo, Spain}
\affiliation[b]{Instituto de Ciencias y Tecnolog\'{\i}as Espaciales de Asturias, c/ Independencia 13, ES-33004 Oviedo, Spain}
\emailAdd{jmbsphysics@outlook.es}
\emailAdd{hoyoscarlos@uniovi.es}
\emailAdd{anton.faedo@uniovi.es}
\emailAdd{mussodaniele@uniovi.es}

\abstract{
We find solutions to consistent truncations of supergravity where some real scalars are analytically extended to imaginary values, ensuring the metric remains real-valued.
Among the solutions there are Lorentzian traversable wormholes connecting two asymptotically anti-de Sitter spacetimes and flows that have a real metric also when uplifted to ten or eleven dimensions.
We argue that the holographic duals are pseudo-Hermitian and $PT$-symmetric theories. Wormhole solutions also admit an interpretation as the low-energy theory of two stacks of branes and anti-branes after tachyon condensation. The wormhole is then dual to an entangled state of two copies of the theory that lives on a stack of branes. We present some evidence by computing the mutual information between the theories at each boundary and by identifying the Goldstone bosons associated with the breaking of the two copies of Poincaré symmetry to their diagonal subgroup.}

\maketitle
\flushbottom
\setcounter{page}{2}

\newpage

%%%%%%%%%%%%%%%%%%%%%%%%%%%%%%%%%%%%%%%%%%%%%%%%%%%%%%%%%
%%%%%%%%%%%%%%%%%%%%%%%%%%%%%%%%%%%%%%%%%%%%%%%%%%%%%%%%%
\section{Introduction}
%%%%%%%%%%%%%%%%%%%%%%%%%%%%%%%%%%%%%%%%%%%%%%%%%%%%%%%%%
%%%%%%%%%%%%%%%%%%%%%%%%%%%%%%%%%%%%%%%%%%%%%%%%%%%%%%%%%

Hermiticity and real-valued couplings are often taken as fundamental requirements in quantum mechanics and quantum field theory (QFT). 
Nevertheless, analytic extensions in coupling space and field space have led to useful insights into a wide range of topics ranging from Chern--Simons theories \cite{Witten:2010cx}, non-perturbative contributions to the path integral \cite{Basar:2013eka} and weak first-order phase transitions \cite{Gorbenko:2018dtm,Gorbenko:2018ncu}, to non-equilibrium physics, open systems, and extensions of quantum mechanics \cite{Mostafazadeh:2008pw,Ashida:2020dkc,Bender:2023cem}.

In the context of the AdS/CFT duality, some analytic extensions of couplings and operators in QFT can, in principle, be realized through the analytic extension of the corresponding fields in the gravitational dual.
This has been previously used to construct bottom-up holographic duals of complex conformal field theories \cite{Faedo:2019nxw,Faedo:2021ksi}, and non-Hermitian, $PT$-symmetric theories \cite{Arean:2019pom,Morales-Tejera:2022hyq,Xian:2023zgu,Arean:2024gks,Arean:2024lzz}. 
Similarly, some wormholes have been interpreted in terms of complex couplings or non-Hermitian density matrices \cite{Maldacena:2018lmt,Loges:2022nuw,Garcia-Garcia:2020ttf,Garcia-Garcia:2022zmo,Harper:2025lav,Kawamoto:2025oko,Held:2026bbo}. In addition, in the discussion of gravitational path integrals and quantum cosmology, analytic extensions of fields are often introduced \cite{Gibbons:1978ac,Halliwell:1989dy,Louko:1995jw}.

The QFT dual to a given geometry is best understood when it can be derived in a UV-complete string-theory framework. We take a step forward in this direction starting from the classical supergravity approximation to string theory.\footnote{See \emph{e.g.} \cite{Nekrasov:2023xzm,Zigdon:2026xal} for discussions of analytic extensions from the sigma model perspective.}  Generically, the analytic extension of a supergravity solution leads to a complex metric, hence a holographic dual QFT that is fully non-unitary with a complex spectrum of conformal dimensions and central charge. Complex metrics may be admissible as saddle points of the path integral under certain conditions 
\cite{Kontsevich:2021dmb,Witten:2021nzp,Lehners:2021mah,Krishna:2026rma}, and complex field theories may contain physically sensible content \cite{Gorbenko:2018dtm,Gorbenko:2018ncu}. However, if the analytic extension is such that the metric remains real, the 
conformal dimensions and central charge will also be real. The situation is similar to that of $PT$-symmetric Hamiltonians. If the ground state is $PT$-symmetric, the energy spectrum is real even if the Hamiltonian is not Hermitian. However, if the ground state breaks $PT$-symmetry, the energies become complex. Thus, the dual of an analytically continued bulk solution with a real metric, if it exists, is expected to be a pseudo-Hermitian theory or a generalization thereof. It is then well motivated to restrict to this subclass of theories with a real-valued metric for the time being, as they are better understood.

In the rest of this paper, we present examples of supersymmetric solutions to consistent truncations of maximal supergravity in four and five dimensions that admit analytic extensions with real metrics. In some of the examples, the uplift to ten and eleven dimensions turns out to also have a real metric and complex fluxes, although this is not true in general. Similar analytic extensions of supergravity have been considered in \cite{Bergshoeff:2007cg,Skenderis:2007sm} in the context of the domain wall/cosmology correspondence \cite{Skenderis:2006jq}. While some solutions are singular, others correspond to wormhole solutions between two asymptotically anti-de Sitter (AdS) spaces, or to flows to an asymptotically flat spacetime.
The supergravity solutions we find differ from the Euclidean axionic wormhole solutions \cite{Giddings:1987cg,Gutperle:2002km,Bergshoeff:2004pg,Bergshoeff:2005zf,Arkani-Hamed:2007cpn,Bergshoeff:2008be,Hertog:2017owm,Ruggeri:2017grz,Marolf:2021kjc,Astesiano:2022qba,Loges:2023ypl,Astesiano:2023iql,Anabalon:2023kcp,DiUbaldo:2026rly}\footnote{See \cite{Hebecker:2018ofv} for a review.} in that we keep the signature fixed and perform the analytic extension only on scalar fields (rather than pseudo-scalars). Similar extensions of scalar fields have been considered recently in a Euclidean setting \cite{Maldacena:2026jqd}.

We study some properties of the wormhole solutions in detail. Despite featuring ghost scalar fields that introduce violations of the null energy condition, two of the three wormholes that we analyze show no linear instability, while the third one has a normal mode with negative mass squared. A possible interpretation of these solutions is that they describe the near-horizon limit of two separated stacks of branes and anti-branes after tachyon condensation has joined them. The low-energy effective theory on the branes describes an entangled state of two copies of the same field theory with imaginary couplings and expectation values. We extract their values applying the holographic dictionary at each boundary and confirm that they are consistent with the dual being a pseudo-Hermitian theory in a state preserving a form of $PT$-symmetry. To further support this interpretation, we compute the mutual information between the degrees of freedom corresponding to the theories at each boundary, and identify a fluctuation corresponding to the Goldstone boson of a broken antidiagonal Poincar\'e symmetry.

The outline of the paper is as follows. In Section \ref{sec:analyticext} we introduce the analytic extension of supergravity solutions leading to real metrics and study several known examples of domain walls. We also study the perturbative stability of wormhole solutions and explain some properties of their fluctuation spectrum. In Section \ref{sec:wormholes} we discuss the possible interpretation of wormhole solutions as the near-horizon limit of brane--anti-brane intersections. In Section \ref{sec:PTsym} we argue that the wormholes are dual to pseudo-Hermitian theories in an entangled state of two copies of a QFT. We support this by computing the entanglement between the two copies and identifying the Goldstone bosons of the broken Poincar\'e symmetries. Finally, in Section \ref{sec:discuss} we discuss the results and possible future directions.

%%%%%%%%%%%%%%%%%%%%%%%%%%%%%%%%%%%%%%%%%%%%%%%%%%%%%%%%%
%%%%%%%%%%%%%%%%%%%%%%%%%%%%%%%%%%%%%%%%%%%%%%%%%%%%%%%%%
\section{Analytic extension of supergravity solutions
}\label{sec:analyticext}
%%%%%%%%%%%%%%%%%%%%%%%%%%%%%%%%%%%%%%%%%%%%%%%%%%%%%%%%%
%%%%%%%%%%%%%%%%%%%%%%%%%%%%%%%%%%%%%%%%%%%%%%%%%%%%%%%%%

String theory at low energies is described by ten-dimensional supergravity, or eleven-dimen\-sional supergravity in its M-theory incarnation. In suitable compactification contexts, it is possible to reduce the theory to a lower-dimensional supergravity and select a subsector of fields that constitutes a consistent truncation.
This means that the solutions to the equations of motion in the consistent truncation are also solutions of the ten- or eleven-dimensional supergravity equations.

We focus on consistent truncations with real scalar fields that are minimally coupled to gravity and have a canonical kinetic term. Denoting the scalars by $\phi^I$ with $I=1,\dots,n$, the supergravity action of this subsector is%
\footnote{We use mostly plus signature for the metric throughout the paper. In our conventions, uppercase Latin indices $M,N,\dots$ label all the directions in the bulk, while Greek indices $\mu,\nu,\dots$ label the directions parallel to the boundary. When directions are only spatial, we use lowercase Latin indices $i,j,\dots$.}
\begin{equation}
    S_{d+1}=\frac{1}{16\pi\, G_{d+1}}\int \dd^{d+1} x\, \sqrt{-G}\left[R-\frac{1}{2}\partial_M \phi^I\partial^M\phi^I-V\left(\{\phi^I\}\right)\right]\ ,
\end{equation}
with $G_{d+1}$ the Newton constant in $d+1$ dimensions. 

If the scalar potential is an even function of one or more scalar fields, it is possible to perform an analytic extension to imaginary values that keeps the action real. Suppose that we order the scalars so that we do the analytic extension on the first $k\leq n$ of them
\begin{equation}
\begin{split}
    &\phi^a\to \phi^a =i\, \eta^a
    \ , \qquad  
    a=1,\dots, k
    \ ,\\
    &\phi^i\to \phi^i
    \ ,\qquad\qquad\quad\,
    i=k+1,\dots,n\ .
\end{split}
\end{equation}
This analytic extension flips the sign of the kinetic term, so the scalars $\eta^a$ become ghosts. In general, this leads to a violation of the energy conditions, including the null energy condition.

If the consistent truncation is supersymmetric, there exists an analytic superpotential $W\left(\{\phi^I\}\right)$ related to the potential through
\begin{equation}
V=\frac{1}{2}\sum_{I=1}^n\left(\frac{\partial W}{\partial \phi^I}\right)^2-\frac{d}{4(d-1)}W^2\ .
\end{equation}
Supersymmetric domain wall solutions can be found using the BPS equations. The metric in domain wall coordinates is of the form
\begin{equation}
    \label{eq:domain_wall}
    \dd s^2
    =
    \dd r^2 + e^{2A(r)}\eta_{\mu\nu}\,\dd x^\mu \dd x^\nu\ ,
\end{equation}
with $\eta_{\mu\nu}$ the $d$-dimensional Minkowski metric. The scalars only depend on the $r$ coordinate. The BPS equations are
\begin{equation}
    \partial_r\phi^I=\partial_{\phi^I}W\ ,\qquad \partial_r A= -\frac{W}{2(d-1)}\ .
\end{equation}
If the superpotential is also an even function of one or several of the $\phi^I$, then the analytic extension of the BPS equations,
\begin{equation}\label{eq:BPS}
    \partial_r \eta^a=-\partial_{\eta^a}\left( W\Big|_{\phi^a=i \eta^a}\right)\ ,\quad \partial_r \phi^i =\partial_{\phi^i}W\Big|_{\phi^a=i \eta^a}\ , \quad  \partial_r A= -\frac{W}{2(d-1)}\Big|_{\phi^a=i \eta^a}\ ,
\end{equation}
gives a supersymmetric domain wall solution with real values for $\eta^a$, $\phi^i$ and $A$. If the ten- or eleven-dimensional uplifted metric is an even function of the analytically continued scalars, then it will also be real. The analytic extension also modifies the potential to
\begin{equation}
\label{eq:VfromW}
V=-\frac{1}{2}\sum_{a=1}^k\left(\frac{\partial W}{\partial \eta^a}\right)^2+\frac{1}{2}\sum_{i=k+1}^n\left(\frac{\partial W}{\partial \phi^i}\right)^2-\frac{d}{4(d-1)}W^2\ .
\end{equation}

A special case where the potential and superpotential are guaranteed to be even functions of a real scalar is when the scalar corresponds to the modulus of a complex field in the larger supergravity theory of which the consistent truncation is a subsector
\begin{equation}
    \Phi^a=\phi^a e^{i\theta^a}\ ,\qquad {\Phi^a}^\dagger=(\Phi^a)^*=\phi^a e^{-i\theta^a}\ .
\end{equation}
The analytic extension can be combined with a shift
\begin{equation}
    \theta^a\to \theta^a=\tilde\theta^a-\frac{\pi}{2}\ .
\end{equation}
Then, the analytic extension of the complex scalar becomes
\begin{equation}
    \Phi^a\to H^a=\eta^a e^{i\tilde\theta^a}\ ,\qquad {\Phi^a}^\dagger\to {H^a}^\dagger=-\eta^a e^{-i\tilde\theta^a}=-(H^a)^*.
\end{equation}

Another possibility is that the internal geometry has a $\mathbb{Z}_2$ symmetry, which translates into a symmetry of the action in the truncated theory under $\Phi^a\to -\Phi^a$. For instance, if the internal geometry is a fibration of two spheres of the same dimension and the scalar corresponds to the ratio of the radii
\begin{equation}
\dd s^2_{\rm int}\sim e^{\alpha}\dd\Omega_n+e^{-\alpha} \dd\widehat \Omega_n\ ,
\end{equation}
the geometry is typically symmetric under $\alpha\to -\alpha$ and the truncated theory including this field depends on even powers of it (or even combinations of exponentials).%
\footnote{
Axions usually come with a $\mathbb{Z}_2$ symmetry too, since they are pseudoscalar fields.
}

%%%%%%%%%%%%%%%%%%%%%%%%%%%%%%%%%%%%%%%%%%%%%%%%%%%%%%%%%
\subsection{Examples of supersymmetric domain wall solutions}
\label{subsec:superpotential}
%%%%%%%%%%%%%%%%%%%%%%%%%%%%%%%%%%%%%%%%%%%%%%%%%%%%%%%%%

We list here a few classic supersymmetric domain wall solutions of consistent truncations that have a superpotential which is an even function of one or several scalar fields. In Appendix~\ref{app:trunc}, a consistent truncation can be found that encompasses most of the five-dimensional models discussed in this section and explicitly shows their charged-scalar origin. 

\begin{enumerate}
\item Solutions dual to the Coulomb branch of ${\cal N}=4$ super Yang--Mills (SYM) preserving $SO(3)\times SO(3)$ symmetry. They can be found using a consistent truncation to $d+1=5$ dimensional supergravity. The superpotential and the solutions, including the uplift to type IIB supergravity, were given in \cite{Freedman:1999gk,Cvetic:1999xx}. There is a single scalar $\phi$ with a superpotential
\begin{equation}
    W(\phi)= \pm \frac{6}{L}\cosh\left(\frac{\phi}{\sqrt{3}}\right)\,.
\end{equation}
The uplift to type IIB supergravity can be interpreted as the geometry produced by a distribution of parallel D3-branes that preserves an $SO(3)\times SO(3)$ subgroup of rotations in the transverse space.

\item Solutions dual to the Coulomb branch of ABJM preserving $SO(4)\times SO(4)$ symmetry. They can be found using a consistent truncation to $d+1=4$ dimensional supergravity. The potential and solutions, including the uplift to eleven-dimensional supergravity, are given in \cite{Cvetic:1999xx}. An explicit expression for the superpotential can be found in \emph{e.g.} \cite{Tarrio:2013qga}. There is a single scalar $\phi$ with a superpotential
\begin{equation}
    W(\phi)=\pm\frac{4}{L}\cosh \left(\frac{\phi}{2}\right)\,.
\end{equation}
The uplift to eleven-dimensional supergravity can be interpreted as the geometry produced by a distribution of parallel M2-branes preserving an $SO(4)\times SO(4)$ subgroup of rotations in the transverse space.

\item Solutions dual to the ${\cal N}=2^*$ SYM theory. They can be found using a consistent truncation to $d+1=5$ dimensional supergravity. The superpotential and the solutions, including the uplift to type IIB supergravity, were given in \cite{Pilch:2000ue}. The extension to a truncation with a complex scalar charged under a gauge field can be found in \cite{Bobev:2010de}. There are two scalar fields, $\chi$ (the modulus of the complex scalar) and $ \alpha$, that can be canonically normalized using $\phi^1=2\sqrt{2}\,\chi$  and $\phi^2= 2\sqrt{6}\,\alpha$. The superpotential for the canonically normalized fields is
\begin{equation}
    W(\phi^1,\phi^2)=\pm\frac{2}{L} e^{-\frac{\phi^2}{\sqrt{6}}} \left[2 + e^{\sqrt{\frac{3}{2}}\phi^2} \cosh\left(\frac{\phi^1}{\sqrt{2}}\right)\right]\ .
\end{equation}
The superpotential is even in $\phi^1$ but not in $\phi^2$, so the analytic extension to purely imaginary values is restricted to $\phi^1$.

\item Solutions dual to the ${\cal N}=1^*$ SYM theory with a mass for one chiral multiplet (the Leigh--Strassler flow). They can be found using a consistent truncation to $d+1=5$ dimensional supergravity. The superpotential and the solutions were given in \cite{Freedman:1999gp,Pilch:2000fu}, with the latter paper also describing the uplift to type IIB supergravity. There are two scalar fields $\chi,\alpha$, that can be canonically normalized using $\phi^1=2\chi$, $\phi^2 =2\sqrt{6}\,\alpha$. The superpotential for the canonically normalized fields has the form
\begin{equation}
    W(\phi^1,\phi^2)= \pm\frac{e^{-\frac{\phi^2}{\sqrt{6}}}}{L}\left\{e^{\sqrt{\frac{3}{2}}\,\phi^2}\left[\cosh\left(\phi^1\right)-3 \right] - 2\left[1 + \cosh\left(\phi^1\right)\right]\right\}\ .
\end{equation}
Also in this case, the superpotential is even in $\phi^1$ but not in $\phi^2$.

\item Solutions dual to the ${\cal N}=1^*$ SYM theory with equal masses for two chiral multiplets (the GPPZ solution). They can be found using a consistent truncation to $d+1=5$ dimensional supergravity. The superpotential and the solutions were given in \cite{Girardello:1999bd}. The complete uplifts to type IIB supergravity were found in \cite{Bobev:2018eer,Petrini:2018pjk}. There are two scalar fields, $m$ and $\sigma$, that can be canonically normalized using $\phi^1= 2m$, $\phi^2 = 2\sigma$. The superpotential for the canonically normalized fields is
\begin{equation}
    W(\phi^1,\phi^2) = \pm \frac{3}{L}\left[\cosh\left({\frac{\phi^1}{\sqrt{3}}}\right)+\cosh\left({\phi^2}\right)\right]\ .
\end{equation}
In this case, both scalars correspond to the modulus of complex fields in a larger truncation, so the superpotential is even in both. 

\item Solutions dual to states with spontaneously broken R-symmetry in $3+1$-dimensional ${\cal N}=1$ SCFTs \cite{Gubser:2009qm}. They can be found using a consistent truncation to $d+1=5$ dimensional supergravity. There is a single canonically normalized scalar, with a superpotential \cite{Cassani:2010uw,Gauntlett:2010vu}
\begin{equation}
    W(\phi)= \pm \frac{6}{L}\cosh^2\left(\frac{\phi}{2}\right)\ .
\end{equation}
\item Solutions dual to states with spontaneously broken R-symmetry in $2+1$-dimensional SCFTs \cite{Gauntlett:2009dn}. They can be found using a consistent truncation to $d+1=4$ dimensional supergravity. There is one scalar field $\chi$, which is related to the canonically normalized field through $\phi=2\tanh^{-1}\frac{\chi}{\sqrt{2}}$. The superpotential for the canonically normalized field is found to be
\begin{equation}
    W(\phi)= \pm \frac{2}{L}\left[\cosh\left(\phi\right)+1\right]=\pm \frac{4}{L}\cosh^2\left(\frac{\phi}{2}\right)\ .
\end{equation}

\item Solutions dual to flows triggered by vevs for boson bilinears. They correspond to the model in Appendix~\ref{app:trunc} with vanishing charged scalars. The canonically normalized neutral fields are $\phi^1=2\sqrt{2}\beta$ and $\phi^2=2\sqrt{6}\alpha$. Despite the fact that $\phi^1$ is not the modulus of a complex scalar, the superpotential is an even function of it 
\begin{align}
W(\phi^1,\phi^2)=\pm\frac{2}{L}e^{-\frac{\phi^2}{\sqrt{6}}}\left[e^{\sqrt{\frac{3}{2}}\phi^2}+2\cosh\left(\frac{\phi^1}{\sqrt{2}}\right)\right]\,.
\end{align}

\end{enumerate}

%%%%%%%%%%%%%%%%%%%%%%%%%%%%%%%%%%%%%%%%%%%%%%%%%%%%%%%%%
\subsection{Analytic extension of domain wall solutions}
\label{subsec:examples}
%%%%%%%%%%%%%%%%%%%%%%%%%%%%%%%%%%%%%%%%%%%%%%%%%%%%%%%%%

The superpotentials we have listed above depend on the scalars through hyperbolic functions. When doing the analytic extension to purely imaginary values, these become trigonometric functions with some periodicity under shifts of the scalars. It is natural to identify critical points of the superpotential as equivalent if they are related by a shift by one period, but there are additional critical points also at half-periods that are inequivalent to the critical point at the origin of the target space.
The flows that are solutions to the BPS equations \eqref{eq:BPS} interpolate between the critical point at the origin of the target space of the scalars and the new critical points of the superpotential. We will use the convention for the sign of the superpotential that puts at $r\to-\infty$ the asymptotic AdS boundary corresponding to the critical point at the origin.

Here we list the solutions we find for each case. 
\begin{enumerate}
    \item The analytic extension $\phi\to \phi=i\, \eta$ makes the superpotential periodic in $\eta$, with period $2\pi\sqrt{3}$,
\begin{equation}
    \label{eq:Wex1}
    W(\eta) = \frac{6}{L} \cos\left(\frac{\eta}{\sqrt{3}}\right)\ .
\end{equation}
There is a flow between the two critical points of the superpotential at $\eta=0$ (when $r\to-\infty$) and $\eta=\pi\sqrt{3}$ (when $r\to +\infty$)
\begin{equation}\label{eq:sol1}
    \eta(r)=2\sqrt{3} \arctan\left(e^{2(r-r_0)/L} \right)
    \ ,\qquad 
    e^{2A(r)}=\frac{e^{2a_0}}{\sin\left(\frac{\eta(r)}{\sqrt{3}}\right)}\ .
\end{equation}
The solution interpolates between two asymptotically $\text{AdS}_{4+1}$ spacetimes of the same radius $L$
\begin{equation}
    e^{2A} \underset{r\to-\infty}{\sim} \frac{e^{2a_0}}{2}e^{-2(r-r_0)/L}
    \ ,\qquad 
    e^{2A} \underset{r\to +\infty}{\sim} \frac{e^{2a_0}}{2}e^{2(r-r_0)/L}\ .
\end{equation}
The warp factor reaches its minimal value at $r=r_0$, $\eta(r_0)=\pi\sqrt{3}/2$, $e^{2A(r_0)}=e^{2a_0}$.

    \item The analytic extension $\phi\to \phi=i\, \eta$ makes the superpotential periodic in $\eta$, with period $4\pi$,
\begin{equation}
    \label{eq:Wex2}
    W(\eta)=\frac{4}{L}\cos\left(\frac{\eta}{2}\right)\ .
\end{equation}
There is a flow between the two critical points of the superpotential at $\eta=0$ (when $r\to-\infty$) and $\eta=2\pi$ (when $r\to +\infty$)
\begin{equation}\label{eq:sol2}
    \eta(r)=4 \arctan\left(e^{(r-r_0)/L} \right)
    \ ,\qquad e^{2A(r)}=\frac{e^{2a_0}}{\sin^2\left(\frac{\eta(r)}{2}\right)}\ .
\end{equation}
The solution interpolates between two asymptotically $\text{AdS}_{3+1}$ spacetimes of the same radius $L$
\begin{equation}
    e^{2A} \underset{r\to-\infty}{\sim} \frac{e^{2a_0}}{4}e^{-2(r-r_0)/L}
    \ ,\qquad 
    e^{2A} \underset{r\to +\infty}{\sim} \frac{e^{2a_0}}{4}e^{2(r-r_0)/L}\ .
\end{equation}
The warp factor reaches its minimal value at $r=r_0$, $\eta(r_0)=\pi$, $e^{2A(r_0)}=e^{2a_0}$.

    \item The analytic extension $\phi^1\to i\eta^1                     $ renders the superpotential periodic in $\eta^1$, with period $2\sqrt{2}\pi$
\begin{equation}\label{eq:supex3}
    W(\eta^1,\phi^2) = \frac{2}{L}e^{-\frac{\phi^2}{\sqrt{6}}} \left[2+e^{\sqrt{\frac{3}{2}}\phi^2}\cos\left(\frac{\eta^1}{\sqrt{2}}\right)\right]\ .
\end{equation}
The solution can be found in implicit form, using $\eta^1$ as radial coordinate,
\begin{subequations}
\begin{eqnarray}
    e^{\sqrt{\frac{3}{2}}\phi^2(\eta^1)} &=&\cos\left(\frac{\eta^1}{\sqrt{2}}\right)+\sin^2\left(\frac{\eta^1}{\sqrt{2}}\right)\left\{C+\log\left[\cot\left(\frac{\eta^1}{2\sqrt{2}}\right)\right]\right\}
    \ ,\\
    e^{2A(\eta^1)} &=& e^{2a_0}\frac{e^{\sqrt{\frac{2}{3}}\phi^2(\eta^1)}}{\sin^2\left(\frac{\eta^1}{\sqrt{2}}\right)}\ .
\end{eqnarray}
\end{subequations}
The solution interpolates between $\eta^1=0$ (for $r\to -\infty$) and a finite value $0<\eta_*<\sqrt{2}\pi$ determined by the integration constant
\begin{equation}
    C=-\log\left[\cot\left(\frac{\eta_*}{2\sqrt{2}}\right)\right]-\frac{\cos\left(\frac{\eta_*}{\sqrt{2}}\right)}{\sin^2\left(\frac{\eta_*}{\sqrt{2}}\right)}\ .
\end{equation}
One has $\eta_* \to 0$ for $C\to -\infty$ and $\eta_* \to \sqrt{2}\pi$ for $C\to +\infty$.

This describes a flow from an asymptotically $\text{AdS}_{4+1}$ space of radius $L$ to a singular geometry
\begin{equation}
    \eta \underset{r\to-\infty}{\sim} \eta_b e^{r/L}
    \ ,\qquad
    e^{2A} \underset{r\to-\infty}{\sim} \frac{2e^{2a_0}}{\eta_b^2}e^{-2r/L}\ ,
\end{equation}
and
\begin{align}
    &\dd s^2\sim \frac{L^2}{2\sin^2\left(\frac{\eta_*}{\sqrt{2}}\right)} e^{-2\sqrt{\frac{2}{3}}\phi^2}\dd\eta^2+\frac{e^{2a_0}}{\sin^2\left(\frac{\eta_*}{\sqrt{2}}\right)}e^{\sqrt{\frac{2}{3}}\phi^2}\eta_{\mu\nu}\dd x^\mu \dd x^\nu
    \ , \\ 
    &\phi^2\sim \sqrt{\frac{2}{3}}\log(\eta_*-\eta) \to -\infty\ .
\end{align}

    \item In this case the domain wall solution has only been found numerically. A preliminary examination indicates that the problem is equally complicated after the analytic extension, so there are no analytic solutions that can be found straightforwardly and we have not attempted to construct numerical solutions in this example.
    
    \item In this case there are several possibilities. We choose to do the analytic extension of both scalar fields $\{\phi^1,\phi^2\}\to \{i\eta^1,i\eta^2\}$. The superpotential is periodic, with periods $2\sqrt{3}\pi$ for $\eta^1$ and $2\pi$ for $\eta^2$:
\begin{equation}
    W(\eta^1,\eta^2) = \frac{3}{L}\left[\cos\left({\frac{\eta^1}{\sqrt{3}}}\right)+\cos\left({\eta^2}\right)\right]\ .
\end{equation}
There is a flow interpolating between two critical points of the superpotential, the origin $(\eta^1,\eta^2)=(0,0)$ (for $r\to -\infty$) and the point $(\eta^1,\eta^2)=(\sqrt{3}\pi,\pi)$ (for $r\to+\infty$) 
\begin{subequations}\label{eq:sol5}
    \begin{equation}
    \eta^1(r) = 2\sqrt{3}\arctan{\left(e^{(r-r_1)/L}\right)}
    \ , \qquad
    \eta^2(r) = 2\arctan{\left(e^{3(r-r_2)/L}\right)}\ ,
\end{equation}
\begin{equation}
    e^{2A(r)}=e^{2a_0}\cosh\left(\frac{r-r_1}{L}\right)\left[\cosh\left(3~\frac{r-r_2}{L}\right)\right]^{1/3}\ .
\end{equation}
\end{subequations}
The solution interpolates between two asymptotically $\text{AdS}_{4+1}$ spacetimes of the same radius $L$
\begin{equation}
    e^{2A} \underset{r\to-\infty}{\sim} \frac{e^{2a_0}}{2^{4/3}}e^{-(2r-r_1-r_2)/L}
    \ ,\qquad 
    e^{2A} \underset{r\to +\infty}{\sim} \frac{e^{2a_0}}{2^{4/3}}e^{(2r-r_1-r_2)/L}\ .
\end{equation}    
    \item  The analytic extension $\phi\to i\eta$ renders the superpotential periodic in $\eta$, with period $2\pi$
\begin{equation}
    W(\eta)= \frac{6}{L}\cos^2\left(\frac{\eta}{2}\right)=\frac{3}{L}\left[1+\cos\left(\eta\right)\right]\ .
\end{equation}
There is a flow between the origin $\eta=0$ (at $r\to-\infty$) and the critical point of the superpotential at $\eta=\pi$ (at $r\to+\infty$)
\begin{equation}
    \eta(r) = 2\arctan{\left(e^{3 (r-r_0)/L}\right)}
    \ ,\qquad 
e^{2A(r)}=e^{2a_0} \left( \sin\frac{\eta(r)}{2}\right)^{-2/3}\ . 
\end{equation}
This flow interpolates between an asymptotically $\text{AdS}_{4+1}$ spacetime and a spacetime that is asymptotically flat
\begin{equation}
    e^{2A} \underset{r\to-\infty}{\sim} e^{2a_0}e^{-2(r-r_0)/L},\quad e^{2A} \underset{r\to +\infty}{\sim} e^{2a_0}\ .
\end{equation}
This solution is particularly interesting in the sense that the uplift to type IIB supergravity \cite{Gubser:2009qm,Cassani:2010uw,Gauntlett:2010vu} retains a real metric, while some of the fluxes supporting it are purely imaginary. It admits any Sasaki--Einstein manifold as internal geometry, including the five-sphere. The metric in ten dimensions reads 
\begin{align}
\dd s^2_{10}=\cos\left(\frac{\eta}{2}\right)\dd s_5^2+\frac{L^2}{\cos\left(\frac{\eta}{2}\right)}\dd s_4^2\left(\text{\small KE}\right)+L^2\cos\left(\frac{\eta}{2}\right)\eta_{\text{\tiny SE}}^2\ ,
\end{align}
where $\dd s_5^2$ is the five-dimensional metric \eqref{eq:domain_wall}, $\dd s_4^2\left(\text{\small KE}\right)$ is the K\"ahler--Einstein base of the Sasaki--Einstein manifold and $\eta_{\text{\tiny SE}}$ describes the ${\rm U(1)}$ fiber over it. This metric is AdS when $\eta\to0$, but is singular in the limit $\eta\to \pi$ in which the five-dimensional metric is simply Minkowski. It is supported by both five- and three-form fluxes, while the axio-dilaton vanishes. In terms of the standard forms defining the Sasaki--Einstein structure (see \cite{Gubser:2009qm,Cassani:2010uw,Gauntlett:2010vu} for details) the two-forms are
\begin{align}
B_2=\im\, L^2\tan\left(\frac{\eta}{2}\right){\rm Re}\,\Omega\,,\qquad\qquad C_2=-\im\,L^2\tan\left(\frac{\eta}{2}\right){\rm Im}\,\Omega\ ,
\end{align}
which are purely imaginary. The self-dual five-form flux, on the other hand, is real and given by
\begin{align}
F_5=L^4\left[2+3\tan^2\left(\frac{\eta}{2}\right)\right]\left(1+*\right)J\wedge J\wedge\eta_{\text{\tiny SE}}\ .
\end{align}

\item The analytic extension $\phi\to i\eta$ renders the superpotential periodic in $\eta$, with period $2\pi$
\begin{equation}
    W(\eta)= \frac{2}{L}\left[\cos(\eta)+1\right]\ .
\end{equation}
There is a flow between the origin $\eta=0$ (at $r\to-\infty$) and the critical point of the superpotential at $\eta=\pi$ (at $r\to+\infty$)
\begin{equation}
    \eta(r) = 2\arctan{\left(e^{2 (r-r_0)/L}\right)}
    \ , \qquad
e^{2A(r)}=e^{2a_0} \left( \sin\frac{\eta(r)}{2}\right)^{-1}\ . 
\end{equation}
This flow interpolates between an asymptotically $\text{AdS}_{3+1}$ spacetime and a spacetime that is asymptotically flat
\begin{equation}
    e^{2A} \underset{r\to-\infty}{\sim} e^{2a_0}e^{-2(r-r_0)/L}
    \ ,\qquad 
    e^{2A} \underset{r\to +\infty}{\sim} e^{2a_0}.
\end{equation}
The solution can be uplifted to eleven dimensions on any Sasaki--Einstein manifold using the results in \cite{Gauntlett:2009dn}. The metric reads
\begin{align}
\dd s^2_{11}=\left[\cos\left(\frac{\eta}{2}\right)\right]^{4/3}\dd s_4^2+\frac{4L^2}{\left[\cos\left(\frac{\eta}{2}\right)\right]^{2/3}}\dd s_6^2\left(\text{\small KE}\right)+4L^2\left[\cos\left(\frac{\eta}{2}\right)\right]^{4/3}\eta_{\text{\tiny SE}}^2\ ,
\end{align}
where $\dd s_4^2$ is the domain-wall metric  \eqref{eq:domain_wall}, $\dd s_6^2\left(\text{\small KE}\right)$ is the metric of a six-dimensional K\"ahler--Einstein base and $\eta_{\text{\tiny SE}}$ is the ${\rm U(1)}$ fiber over it. There is an IR singularity as $\eta\to\pi$. This metric is supported by the four-form flux  
\begin{align}
F_4&=\frac{3}{L}\left[1+\frac43\tan^2\left(\frac{\eta}{2}\right)\right]\cos^4\left(\frac{\eta}{2}\right){\rm vol}_4-32\im L^3\tan\left(\frac{\eta}{2}\right){\rm Re}\,\Omega\wedge\eta_{\text{\tiny SE}}\nonumber\\[2mm]
&+16\im L^2\tan\left(\frac{\eta}{2}\right)\dd r\wedge{\rm Im}\,\Omega\ ,
\end{align}
with ${\rm vol}_4$ the volume form of the four-dimensional metric \eqref{eq:domain_wall}. The definition and properties of the one-form $\eta_{\text{\tiny SE}}$ and complex three-form $\Omega$ can be found in \cite{Gauntlett:2009dn}. Notice that the internal part is purely imaginary, while the spacetime components are real. 

\item Performing the analytic extension $\phi^1\to\im\eta^1$ while keeping $\phi^2$ real results in a periodic superpotential with period $2\sqrt{2}\pi$
\begin{align}
W(\eta^1,\phi^2)=\frac{2}{L}e^{-\frac{\phi^2}{\sqrt{6}}}\left[e^{\sqrt{\frac{3}{2}}\phi^2}+2\cos\left(\frac{\eta^1}{\sqrt{2}}\right)\right]\,.
\end{align}
The structure of this superpotential is very similar to that in \eqref{eq:supex3}. The solution can be given again using $\eta^1$ as radial coordinate
\begin{align}
    e^{2A(\eta^1)}=e^{2a_0}\frac{e^{-\frac{\phi^2(\eta^1)}{\sqrt{6}}}}{\sin\left(\frac{\eta^1}{\sqrt{2}}\right)}
    \ ,\qquad e^{\sqrt{\frac32}\phi^2(\eta^1)}=\frac{1}{\cos\left(\frac{\eta^1}{\sqrt{2}}\right)+C\sin\left(\frac{\eta^1}{\sqrt{2}}\right)}\,.
\end{align}
The radial coordinate ranges between $\eta^1=0$ and $\eta^1=\eta_*$, where the upper value depends on the integration constant as $C=-\cot\left(\eta_*/\sqrt{2}\right)$. It interpolates between ${\rm AdS}_{4+1}$ and a hyperscaling-violating geometry with coefficient $\theta=4$.

\end{enumerate}

The solutions in Examples 1, 2 and 5, interpolating between two asymptotically AdS spaces, can be thought of as instances of ``traversable wormholes'' where the only way to travel between the asymptotic regions is through the wormhole itself.  Solutions 1 and 2 turn out to be particular examples of the families studied in \cite{Huang:2020qmn,Nozawa:2020gzz}.

Without a violation of the null energy condition, this type of solution should not be possible \cite{Morris:1988tu,Hochberg:1998ii}. Nonetheless, such violation does not lead to causality violations of the type discussed in \cite{Morris:1988tu}, since it is not possible to travel between the two asymptotic regions outside the wormhole. There are also no causality violations for signals connecting two points at the boundary, of the type discussed in \cite{Hoyos:2010at}, as the local speed of light is the same throughout the geometry, this latter being manifest from the domain-wall ansatz \eqref{eq:domain_wall}. In the rest of the paper we will focus on these solutions, as they are the most interesting for their connection to the holographic description of entanglement and to the quantum gravitational path integral.

%%%%%%%%%%%%%%%%%%%%%%%%%%%%%%%%%%%%%%%%%%%%%%%%%%%%%%%%%
\subsection{Perturbative stability and fluctuation spectrum of wormholes}
%%%%%%%%%%%%%%%%%%%%%%%%%%%%%%%%%%%%%%%%%%%%%%%%%%%%%%%%%

Our starting point have been supergravity theories, for which there are positive energy proofs that ensure stability of supersymmetric solutions \cite{Gibbons:1983aq,Hull:1983ap,Townsend:1984iu}. The solutions we are constructing satisfy a set of BPS equations and can be seen as analytic extensions of supersymmetric solutions. One might wonder if the stability proofs based on supersymmetry could also be generalized. However, while the positive energy theorems allow for scalar potentials that are unbounded from below, they still require kinetic terms with the usual sign,\footnote{This is more clearly seen in \cite{Townsend:1984iu}.} so they do not extend straightforwardly to the examples considered here. 

In this subsection, we study the perturbative stability of backgrounds with only ghost scalars coupled to gravity. More complicated cases with both ghost and ordinary scalars are left for future studies.

We expand around the background domain wall solution to linear order in the metric and scalar fluctuations. We enforce the radial gauge and consider a plane wave expansion. We then identify the combinations of fluctuating fields that are invariant under the remaining diffeomorphisms that are not fixed by the gauge choice. Full details can be found in Appendix \ref{app:eoms}. For invariant fluctuations involving only the metric components, the equations of motion take the form derived in \eqref{eq:invariantHeq}, in $d+1$ dimensions,
\begin{equation}
    H''+(d-4)A' H'+\left[e^{-2A}M^2-2(d-2)(A')^2-\frac{1}{d-1}(\eta_0')^2\right]H=0\ ,
\end{equation}
with $H$ the invariant fluctuation and $M^2=-\eta^{\mu\nu}k_\mu k_\nu$ the mass for a mode with momentum $k_\mu$ along the field theory directions. The radial functions $\eta_0(r)$ and $A(r)$ are the background values of the ghost scalar and the warp factor, respectively. 

It is convenient to write the equation in Sturm--Liouville form
\begin{equation}
    \frac{d}{dr}\Big[p(r)H'(r)\Big]+q(r)H(r)+M^2 w(r) H(r)=0\ ,
\end{equation}
with
\begin{equation}
    p(r)=e^{(d-4)A}
    ,\quad 
    q(r)=-e^{(d-4)A}\left[2(d-2)(A')^2+\frac{1}{d-1}(\eta_0')^2\right],\quad w(r)=e^{(d-6)A}.
\end{equation}
Normal modes are solutions to the equations that are normalizable at the asymptotic boundaries (in the case of wormholes), or normalizable at the asymptotic boundary and regular in the interior. This is a Sturm--Liouville problem with eigenvalues $\lambda=M^2$. Non-trivial solutions will typically exist only for a discrete set of values $\lambda_n=M_n^2$, $n=1,2,3,\dots$. Since these are real equations with real coefficients and the boundary conditions are also real, the modes and eigenvalues will also be real. Therefore, an instability will happen iff there is an eigenmode with a negative eigenvalue $\lambda_n=M_n^2<0$. The modes in the spectrum split according to parity in the holographic radial direction $r-r_0\to -(r-r_0)$, if the geometry has this symmetry.

Now we show that all eigenvalues are positive. For this, we multiply the equation in Sturm--Liouville form by the eigenfunction and integrate over the radial coordinate. We find, for any $n$,
\begin{equation}
    \int_a^b  \dd r \, \left[H_n\frac{d}{dr}\Big(p H_n'\Big)+q H_n^2+M_n^2 w H_n^2\right]=0\ ,
\end{equation}
where the integration domain $(a,b)$ spans the entire range of the radial coordinate. We can integrate by parts the first term inside the integral. For a normal mode, the resulting boundary term should vanish at the extremes of the interval, so after moving some terms to the other side of the equality, we are left with
\begin{equation}
    M_n^2\int_a^b  \dd r \, w H_n^2=\int_a^b  \dd r \, \left[p (H_n')^2-q H_n^2\right]\ .
\end{equation}
Since $H_n$ is real, and $p>0$, $q<0$, and $w>0$ in the full range of the integration, the integrals on both sides of the equality are positive definite and it follows that $M_n^2>0$.
This proves that these backgrounds are perturbatively stable as far as the $H$ fluctuations are concerned. Curiously, this argument does not work for an ordinary scalar, since the term in the coefficient $q$ proportional to $(\eta_0')^2$ would have the opposite sign, and $q$ would not have a definite sign.

To complete the stability analysis, we need to study the remaining gauge-invariant fluctuation $Z_\eta$, the one which involves the ghost scalar. We define it in \eqref{eq:zetaeta} and its equation of motion is derived in \eqref{eq:invariantZeq},
\begin{equation}
    \label{eq:eom_Zeta}
    0=Z_\eta''+dA' Z_\eta'+\left(e^{-2A}M^2+Q_{ZZ}\right)Z_\eta+Q_{ZH}(H_S'-2A' H_S).
\end{equation}
We remind the reader that $H_S$ is a gauge-invariant, scalar mode that involves only the metric fluctuations \eqref{eq:HS}. 

One can study the spectrum of \eqref{eq:eom_Zeta} setting $H_S=0$. In fact, $Z_\eta$ does not source other modes and we have already determined that the spectrum of $H_S$ is stable. Unfortunately, the coefficient $Q_{ZZ}$ given in \eqref{eq:Qcoefs} and \eqref{eq:Qcoef2} does not have a definite sign, so the Sturm--Liouville argument that we used above does not apply and it is necessary to compute the spectrum of $Z_\eta$ explicitly. We have performed the calculation of the first eigenvalues in the spectrum for Examples 1, 2, and 5\footnote{Note that Example 5 features two scalar ghosts, but the analysis is similar to the cases with just a single scalar ghost; the metric and ghost fluctuation spectra can be studied separately.} from Subsection \ref{subsec:examples}. For convenience, we have adopted different approaches, Chebyshev spectral method, analytic analysis, and numerical shooting, respectively. In Examples 1 and 2, with a single ghost scalar, we found no evidence of instability. Example 5, with two coupled scalars, features an unstable mode (see Appendix \ref{app:numerics} for further details).

It might be puzzling that a theory containing a field with the `wrong' sign of the kinetic term could be stable, as we find in Examples 1 and 2. There are, however, some heuristic arguments that might help to understand the stability of the solutions. Neglecting the coupling to gravity, the ghost scalar has a Hamiltonian density that is non-positive and thus bounded from above. Having any conserved quantity bounded either from above or from below makes a system with ghosts stable \cite{ErrastiDiez:2024hfq}. When one considers the coupling to gravity, having a more negative energy density from the ghost kinetic term may actually help to stabilize the system. Consider the case where only a constant potential contributes. A negative potential coupled to gravity results in an AdS solution, which allows for scalar fields with negative mass squared above the Breitenlohner--Freedman bound. The more negative the potential, the more negative the mass squared of the fields can be, thus the AdS space is in some sense more stable under linearized perturbations as the energy density becomes more negative. Although the contribution from a kinetic term is not exactly the same as that of the potential, for the ghost fields the kinetic terms add to the potential to make the energy density even more negative.

%%%%%%%%%%%%%%%%%%%%%%%%%%%%%%%%%%%%%%%%%%%%%%%%%%%%%%%%%
%%%%%%%%%%%%%%%%%%%%%%%%%%%%%%%%%%%%%%%%%%%%%%%%%%%%%%%%%
\section{Wormholes and tachyon condensation}\label{sec:wormholes}
%%%%%%%%%%%%%%%%%%%%%%%%%%%%%%%%%%%%%%%%%%%%%%%%%%%%%%%%%
%%%%%%%%%%%%%%%%%%%%%%%%%%%%%%%%%%%%%%%%%%%%%%%%%%%%%%%%%

The wormhole solutions connect two asymptotic AdS regions of the same radius, which suggests that there are two quantum field theories with the same number of degrees of freedom that are decoupled from each other in the $N\to \infty$ limit, where the gravitational dual becomes free. The main difference that one observes between the two regions connected through the wormhole is that the orientation of the boundary is reversed, which is also reflected in the superpotential having an opposite sign in each asymptotic region. When the AdS solutions at the critical points of the superpotential have a dual interpretation in terms of a brane intersection, this suggests that each boundary would correspond to an opposite orientation of the branes and thus one boundary would correspond to a stack of branes and the other to a stack of anti-branes. 
 
\begin{figure}[t!]
    \centering

    \includegraphics[width=0.7\linewidth]{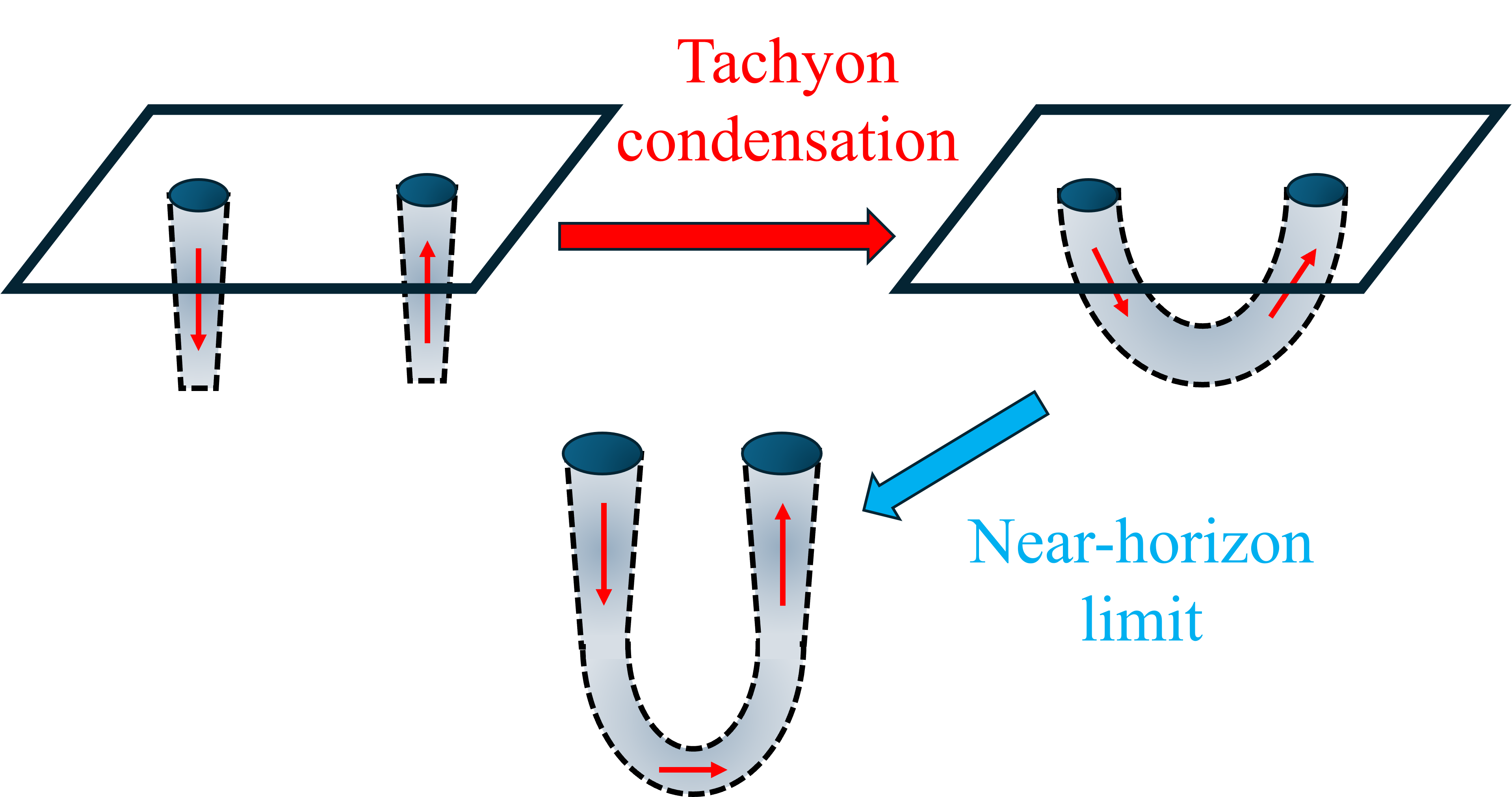}
    \caption{Possible formation of the wormhole. Two stacks of branes and anti-branes form throats inside an asymptotically flat spacetime when they backreact. Condensation of the open string tachyon makes the two throats connect forming a wormhole. In the near-horizon limit the geometry is restricted to the interior of the wormhole, which connects two asymptotically AdS spacetimes.}
    \label{fig:tachyon}
\end{figure}

The observations made above hint at the possibility that the wormhole solution is dual to a state where tachyon condensation connects branes with anti-branes. An analogous non-gravitational example is the description of chiral symmetry breaking in the Sakai--Sugimoto model when D8-branes and anti-branes reconnect in the gravity dual \cite{Sakai:2004cn,Casero:2007ae,Dhar:2007bz}. With this interpretation, the geometry of the wormhole could be as described in Figure \ref{fig:tachyon}, similar to the one discussed in \cite{Balasubramanian:2020ffd} for two stacks of D3-branes. One starts with an asymptotically flat geometry and two separated throats corresponding to each of the stacks. Then the throats become connected at the interior by cutting and gluing them, and the asymptotically flat region is removed by taking the near-horizon limit. Nevertheless, the solutions discussed here present two differences with respect to the construction of \cite{Balasubramanian:2020ffd}: first, the adhesion of the throats does not require an inversion; second, the near-horizon limit here would be such that only the low-energy degrees of freedom living in the wormhole remain, and the two throats do not join in a larger throat with a single AdS boundary. Note that, before taking the near-horizon limit, there would be an attractive force between the two throats that renders the system unstable. However, the near-horizon limit removes the degrees of freedom corresponding to the center of mass motion of each of the throats and this instability would disappear, although there could be other instabilities localized inside the throat.

Let us provide a concrete example that exhibits some of the characteristics mentioned above. We do not claim that this example is the actual holographic dual of the flows we found, but it might serve to illustrate the connection between the analytic extension of the fields and tachyon condensation.

An effective action for a system of D-branes and anti-D-branes was proposed in \cite{Sen:2003tm,Garousi:2004rd} (see also \cite{Kraus:2000nj,Takayanagi:2000rz,Jones:2002sia,Jatkar:2000ei,Hikida:2000cp}). It takes the form of a generalization of the Dirac--Born--Infeld and Wess--Zumino actions including the open string tachyon. For D$p$-branes, the low-energy expansion results in two copies of the maximally supersymmetric theory in $p+1$ dimensions plus a complex tachyon field, with an action 
\begin{equation}
    S_{D\bar{D}}=S_{\text{SYM}}^{(1)}+S_{\text{SYM}}^{(2)}+S_{\rm tach}\ ,
\end{equation}
where, assuming an Abelian configuration of $N$ $D\bar{D}$ pairs,
\begin{equation}
    S_{\rm tach}=-N(2\pi \alpha')^2T_{Dp}\int \dd^{p+1}x \left(\partial_\mu\overline{T}\partial^\mu T+\frac{1}{4}\partial_\mu X\partial^\mu X+V\right),
\end{equation}
where $T$ is the tachyon and $X=X^{(1)}-X^{(2)}$ is the separation between the stacks of branes and anti-branes. The potential is%
\footnote{Before expanding in $\alpha'$, the potential decays exponentially at large values of the tachyon to the true minimum at $T\to \infty$. This behavior is lost in the low-energy expansion.}
\begin{equation}
    V(|T|,X)=|T|^2\left( X^2-\frac{1}{2\alpha'}\right)+v_4 (|T|^2)^2+V_0\ .
\end{equation}
The first two terms correspond to the mass of the lowest energy excitation of a string extended between the brane and the anti-brane. The quartic term depends on details of the tachyon potential which are not fully known, but, from the proposals that have been put forward, one expects to have $v_4>0$ and of order one. A physical separation between the two stacks of the order of the size of the throats, $L$, corresponds to $X^2\sim L^2/(\alpha')^2\gg 1/\alpha'$ when the number of branes is large enough. For instance, in the case of D3-branes, $L$ gives the AdS radius of the near-horizon geometry, and it is related through the holographic dictionary to the 't Hooft coupling of the field theory $X^2\sim \sqrt{\lambda}/\alpha'\gg 1/\alpha'$. In this case, the tachyon is stabilized at $T=0$ and there is no condensation; the branes and anti-branes are completely decoupled at low energies. However, if we now allow the analytic extension to purely imaginary values of the separation, $X\to X=i Y$, the $T=0$ point becomes a maximum and tachyon condensation is triggered. The wormhole would be the endpoint of the condensation in the holographic dual description.

%%%%%%%%%%%%%%%%%%%%%%%%%%%%%%%%%%%%%%%%%%%%%%%%%%%%%%%%%
%%%%%%%%%%%%%%%%%%%%%%%%%%%%%%%%%%%%%%%%%%%%%%%%%%%%%%%%%
\section{Holographic interpretation of wormhole solutions}\label{sec:PTsym}
%%%%%%%%%%%%%%%%%%%%%%%%%%%%%%%%%%%%%%%%%%%%%%%%%%%%%%%%%
%%%%%%%%%%%%%%%%%%%%%%%%%%%%%%%%%%%%%%%%%%%%%%%%%%%%%%%%%

For wormhole solutions connecting two asymptotic AdS regions, the interpretation is that there is a Hilbert space associated with each of the boundaries and that the wormhole is an entangled state in the tensor product of the two Hilbert spaces. For an eternal black hole with asymptotic regions connected by an Einstein--Rosen bridge, the state is identified as the thermofield double state \cite{Maldacena:2001kr}. The ER bridge is not a traversable wormhole, but it can be made traversable by introducing double-trace couplings between the two boundaries \cite{Gao:2016bin}. For non-Hermitian systems, interactions may not be necessary in order to have a traversable wormhole \cite{Harper:2025lav}.

%%%%%%%%%%%%%%%%%%%%%%%%%%%%%%%%%%%%%%%%%%%%%%%%%%%%%%%%%
\subsection{Pseudo-Hermiticity and $PT$-symmetry}
%%%%%%%%%%%%%%%%%%%%%%%%%%%%%%%%%%%%%%%%%%%%%%%%%%%%%%%%%
%

In our setup we have two non-interacting copies of deformed super Yang--Mills theories. We extend analytically some of the operators and couplings in such a way that the theory is pseudo-Hermitian and $PT$-symmetric for appropriately defined transformations. 

Let us consider a set of gauge-invariant operators in a representation $R$ (and its complex conjugate $\overline{R}$) of the global symmetry group, and their associated couplings
\begin{equation}
    {\cal O}_R\ ,\qquad g_R\ ;\qquad {\cal O}_{\overline{R}}\ ,\qquad g_{\overline{R}}\ .
\end{equation}
The quantum effective action of the whole system is the sum of the actions of each of the copies, one holomorphically extended to complex values and the other anti-holomorphically extended
\begin{equation}
    \Gamma=\Gamma_{SYM}\left[{\cal O}_R^{(1)},g_R=g\right]+\Gamma_{SYM}\left[\overline{{\cal O}}_{\overline{R}}^{(2)},g_{\overline{R}}=\overline{g}\right]\ .
\end{equation}
The system is pseudo-Hermitian, since there is an invertible transformation $W$ such that
\begin{equation}
\Gamma^\dagger=    W^{-1} \Gamma W\ .
\end{equation}
The transformation is\footnote{The transformation involves a transposition of the symmetry generators that changes the representation to its conjugate.}
\begin{equation}
    W: {\cal O}^{(1)}_R\to {\cal O}^{(2)}_{\overline{R}},\quad {\cal O}^{(2)}_R\to {\cal O}^{(1)}_{\overline{R}}\ .
\end{equation}
We introduce now the anti-linear transformation $T$ implementing complex conjugation
\begin{equation}
    T: {\cal O}^{(1)}_R\to \overline{{\cal O}}^{(1)}_{\overline{R}} ,\quad  {\cal O}^{(2)}_R\to \overline{{\cal O}}^{(2)}_{\overline{R}} ,\quad g\to \overline{g}\ ,
\end{equation}
and the linear transformation
\begin{equation}
    P: {\cal O}^{(1)}_R\to {\cal O}^{(2)}_R,\quad {\cal O}^{(2)}_R\to {\cal O}^{(1)}_R\ .
\end{equation}
Together, they combine in the $PT$ transformation
\begin{equation}
    PT: {\cal O}^{(1)}_R\to \overline{{\cal O}}^{(2)}_{\overline{R}}, \quad {\cal O}^{(2)}_R\to \overline{{\cal O}}^{(1)}_{\overline{R}},\quad g\to \overline{g}\ ,
\end{equation}
that exchanges the operators of the two copies and leaves the effective action invariant:
\begin{equation}
   (PT)^{-1}\, \Gamma\, PT=\Gamma\ .
\end{equation}
The expectation values of operators in a $PT$-invariant state are related as follows
\begin{equation}
    \vev{\cO_R^{(1)}}=\bra{\Omega} \cO_R^{(1)} \ket{\Omega}=\bra{\Omega} (PT)^{-1}\cO_R^{(1)} PT\ket{\Omega}=\bra{\Omega} \overline{\cO}_{\overline{R}}^{(2)} \ket{\Omega}=\vev{\overline{\cO}_{\overline{R}}^{(2)}}=\vev{\cO_R^{(2)}}^*\ .
\end{equation}
For purely imaginary values, this implies $\vev{\cO_R^{(2)}}=-\vev{\cO_R^{(1)}}$.

In the holographic theory, we can obtain the expectation value of a scalar operator of scaling dimension $\Delta$ from the canonical momentum of the dual scalar field. The asymptotic expansion at each of the boundaries is
\begin{equation}\label{eq:asymptexp}
    \phi\underset{r\to \mp\infty}{\sim} \phi_\mp+\alpha_\mp e^{\pm (d-\Delta) r/L}+\beta_\mp e^{\pm \Delta r/L}\ ,
\end{equation}
where $\phi_\pm$ is the constant asymptotic value of the scalar. In ordinary quantization the couplings $g$ for the dual operators in each copy of the field theory are proportional to the coefficients of the leading terms, $\alpha_\pm$, while the expectation values are determined by the coefficients of the subleading terms $\beta_\pm$. In the alternative quantization, the role of the coefficients is exchanged.

In ordinary quantization, $\Delta>d-\Delta$,
\begin{equation}\label{eq:vevordquant}
\begin{split}
    \vev{\cO^{(1)}}=&\ \frac{L^{d-\Delta}}{2\kappa^2_{d+1}}\lim_{r\to -\infty} e^{(d-\Delta)r/L}e^{dA}\left(\partial_r\phi-\partial_\phi W\right)\ ,\\
    \vev{\cO^{(2)}}=&\ -\frac{L^{d-\Delta}}{2\kappa^2_{d+1}}\lim_{r\to +\infty} e^{-(d-\Delta)r/L}e^{dA}\left(\partial_r\phi-\partial_\phi W\right)\ .
\end{split}
\end{equation}
The terms depending on the superpotential originate from counterterms that are introduced following the usual holographic renormalization prescription at each asymptotic boundary. Note that for BPS flows in ordinary quantization the expectation value is zero, but the formulas are valid only when the coefficient of the quadratic term in the superpotential is proportional to $d-\Delta\leq \Delta$, otherwise the leading divergences are not cancelled. For BPS flows with $\alpha_\pm=0$ the superpotential used for renormalization is different in general from the superpotential in the BPS equations, and one might find non-zero values.

The analytic extension to imaginary values $\phi\to i\eta$ gives
\begin{equation}\label{eq:vevordquantIm}
\begin{split}
    \vev{\cO^{(1)}}=&\frac{iL^{d-\Delta}}{2\kappa^2_{d+1}}\lim_{r\to -\infty} e^{(d-\Delta)r/L}e^{dA}\left(\partial_r\eta+\partial_\eta W\right),\\
    \vev{\cO^{(2)}}=&-\frac{iL^{d-\Delta}}{2\kappa^2_{d+1}}\lim_{r\to +\infty} e^{-(d-\Delta)r/L}e^{dA}\left(\partial_r\eta+\partial_\eta W\right).
\end{split}
\end{equation}
The asymptotic expansion of $i\eta$ is the same as \eqref{eq:asymptexp}, with purely imaginary values $\phi_\pm=i \eta_\pm$, $\alpha_\pm=i \tilde\alpha_\pm$, $\beta_\pm=i\tilde\beta_\pm$.

For the flows that are not wormholes, but end at a singularity or flow to asymptotically flat spacetime, the $PT$ transformation exchanges solutions with the opposite sign of the superpotential. This reverses the sign of the purely imaginary expectation values of the scalars. The previous discussion would still apply to those cases, except that the two copies of the theory would be in a state that factorizes.

%%%%%%%%%%%%%%%%%%%%%%%%%%%%%%%%%%%%%%%%%%%%%%%%%%%%%%%%%
\subsubsection{Examples}
%%%%%%%%%%%%%%%%%%%%%%%%%%%%%%%%%%%%%%%%%%%%%%%%%%%%%%%%%

\begin{enumerate}
    \item This is a special case where $\Delta=d-\Delta=2$, so the superpotential that cancels the divergences has the same form as the BPS superpotential, with the freedom of adding an additional finite counterterm. In the absence of sources, this implies that the expectation values vanish, so the $PT$-symmetry condition is trivially satisfied 
    \begin{equation}     
    \vev{\cO^{(2)}}=-\vev{\cO^{(1)}}=0\ .
    \end{equation}
    \item In this case the field theory lives in $d=3$ dimensions. In ordinary quantization the dimension of the scalar operator would be $\Delta=2$, and the expectation values according to \eqref{eq:vevordquantIm} would be
\begin{equation}
    \vev{\cO^{(1)}}_{\Delta=2}=-\frac{i}{2\kappa_4^2} \tilde\beta_-
    \ ,\qquad
    \vev{\cO^{(2)}}_{\Delta=2}=-\frac{i}{2\kappa_4^2} \tilde\beta_+\ .
\end{equation}
However, in this case one must use alternative quantization in such a way that the dimension of the dual operator corresponds to a scalar bilinear with $\Delta=1$. This is achieved by adding a term  $\Delta {\cal L}=c\,\alpha \beta$ in the boundary action that cancels the variation with respect to the leading coefficient. Denoting $\Delta_1$ as the dimension of the operator in ordinary quantization,
\begin{equation}
\begin{split}
    \delta S_{\text{SUGRA}}=&\int \dd^dx\left(L^{\Delta_1-d}\vev{\cO}_{\Delta_1}\delta \alpha+c\delta(\alpha \beta)\right)\\
    &=\int \dd^d x\left(0 \delta \alpha+c\,\alpha\delta \beta\right)=\int \dd^d x L^{-\Delta_1}\vev{\cO}_{d-\Delta_1}\delta \beta\ .
\end{split}
\end{equation}
The expectation values are
\begin{equation}\label{eq:vevaltquant}
    \vev{\cO^{(1)}}_{\Delta=1}=\frac{i L}{2\kappa_4^2 } \tilde\alpha_-
    \ ,\qquad
    \vev{\cO^{(2)}}_{\Delta=1}=\frac{iL}{2\kappa_4^2} \tilde\alpha_+\ .
\end{equation}
The expansions of \eqref{eq:sol2} in the symmetric case $r_0=0$ give the coefficients
\begin{equation}
\begin{split}
   & \eta_-=0,\ \tilde\alpha_-=4,\ \tilde\beta_-=0\ ,\\
   & \eta_+=2\pi,\ \tilde\alpha_+=-4,\ \tilde\beta_+=0\ .
\end{split}
\end{equation}
We see that the result is consistent with the symmetric wormhole being dual to a $PT$-symmetric state
\begin{equation}
    \vev{\cO^{(2)}}_{\Delta=1}=-\vev{\cO^{(1)}}_{\Delta=1}=-\frac{2iL}{\kappa_4^2}\ .
\end{equation}

\item[5.] The solution for $\eta^1$ in \eqref{eq:sol5} corresponds to introducing a mass deformation, while the solution for $\eta^2$ corresponds to an expectation value. Since the solutions are BPS, the expectation value of the operator dual to $\eta^1$ vanishes. We find that the couplings and the expectation values for each of the operators satisfy the $PT$-symmetric relations
\begin{equation}
    \tilde\alpha_-^1=-\tilde\alpha_+^1=2\sqrt{3}
    \ ,\qquad
    \vev{\cO_2^{(2)}}=-\vev{\cO_2^{(1)}}=-\frac{2i}{\kappa^2_{5}}\ .
\end{equation}

\end{enumerate}

Note that a double-trace deformation coupling operators in each of the two copies
\begin{equation}
    \Delta {\cal S}_{DT}=\int \dd^dx\,\lambda_{DT} \, {\cal O}^{(1)}(x){\cal O}^{(2)}(x)\ ,
\end{equation}
would fix a relation between the expectation value of the scalar operator at one of the boundaries of the wormhole and the source at the opposite boundary, following the usual arguments \cite{Witten:2001ua}. Since the expectation values are non-zero but the sources vanish, we can conclude that there are no double-trace deformations coupling the two copies of the scalar operator, which in other setups are needed to make wormholes traversable \emph{e.g.} \cite{Gao:2016bin,Maldacena:2017axo,Maldacena:2018lmt,Caceres:2018ehr,Bak:2018txn,Ahn:2022suv,Khairunnisa:2025ljk,Djogama:2026pmd}.

%%%%%%%%%%%%%%%%%%%%%%%%%%%%%%%%%%%%%%%%%%%%%%%%%%%%%%%%%
\subsection{Entanglement between the two QFTs}
%%%%%%%%%%%%%%%%%%%%%%%%%%%%%%%%%%%%%%%%%%%%%%%%%%%%%%%%%

We can give a measure of the entanglement of the state dual to the wormhole by computing the mutual information between the degrees of freedom associated with each of the asymptotic AdS boundaries inside some region. 

Let us take a strip bounded by two planes separated in the $x$ direction, and compactify the transverse spatial directions on a torus $T^{d-2}$. The entanglement entropy of the strip 
can be computed following the Ryu--Takayanagi prescription as the area of the minimal surface attached to the boundaries of a strip at each of the AdS boundaries. The full entanglement entropy of the strip taking into account all degrees of freedom corresponds to a minimal surface that is attached to both boundaries. In this case, there are two possibilities: either the minimal surface consists of two disconnected surfaces each attached to one of the boundaries, or the minimal surface connects the two boundaries through the wormhole. For the former case, the mutual information is zero, while for the latter it can be non-zero.

We will denote the area (in Planck units) of a minimal surface attached to strip A at a single boundary as $S_d(\text{A})$, and the area of the minimal surface connecting the same side of the strip at the two boundaries through the wormhole as $S_c(\text{A})$. The mutual information is
\begin{equation}
    I(\text{A})={\rm max}\,\Big\{2 \left[S_d(\text{A})-S_c(\text{A})\right],0\Big\}\ .
\end{equation}

We now proceed to compute the areas. The surface is parameterized by a profile function $x(r)$, in such a way that the induced metric on a constant time slice is
\begin{equation}
    \dd s^2
    =
    \left[1+e^{2A}(x')^2\right]\, \dd r^2+e^{2A}\dd s^2_{T^{d-2}}\ .
\end{equation}
The area in Planck units is
\begin{equation}
    S=\frac{V(T^{d-2})}{4 G_{d+1}}\int \dd r\, e^{(d-2)A}\sqrt{1+e^{2A}(x')^2}\ .
\end{equation}
Using that translations in $x$ are a symmetry, so the conjugate momentum $P$ is constant, one obtains the following equation
\begin{equation}\label{eq:solprofile}
    x'=\pm P\frac{e^{-A}}{\sqrt{e^{2(d-1)A}-P^2}}\ .
\end{equation}
Evaluating this on the area gives
\begin{equation}
S= \frac{V(T^{d-2})}{4 G_{d+1}}\int \dd r\, \frac{e^{(2d-3)A}}{\sqrt{e^{2(d-1)A}-P^2}}\ .
\end{equation}
Note that the warp factor $e^A$ reaches a minimal value at some $r=r_0$ (if the wormhole is symmetric, $r_0=0$). If $P^2< e^{2(d-1)A(r_0)}$, then the argument of the square root never vanishes and there is a profile that extends across the wormhole. This corresponds to the surfaces connecting the two boundaries. On the other hand, if $P^2> e^{2(d-1)A(r_0)}$ the profile stops at some $r_*\neq r_0$ and each of the two branches in \eqref{eq:solprofile} describes one half of a minimal surface ending at a single boundary.

We are taking the same strip on both boundaries; this means that the surfaces connecting the same side of the strip on the two boundaries should have $P=0$ for any width of the strip. On the other hand, for surfaces that are attached at a single boundary, we can characterize the profile by the value closest to the wormhole's throat that can be reached, $r=r_*$, by setting $P=e^{(d-1)A(r_*)}$. Assuming a symmetric wormhole, $r_0=0$ and $A(-r)=A(r)$, we arrive at
\begin{align}\label{eq:mutualinfstrip}
    &2 \left(S_d(\text{A})-S_c(\text{A})\right)
    =\\ &\qquad \nonumber
    \frac{V(T^{d-2})}{G_{d+1}}\left\{\int_{r_*}^\infty \dd r\, \left[\frac{e^{(2d-3)A}}{\sqrt{e^{2(d-1)A}-e^{2(d-1)A(r_*)}}}-e^{(d-2)A}\right]-\int_0^{r_*}\dd r e^{(d-2)A} \right\}\ .
\end{align}
The width of the strip is
\begin{equation}\label{eq:widthstrip}
    \Delta \ell(\text{A})=2e^{(d-1)A(r_*)}\int_{r_*}^\infty \dd r\, \frac{e^{-A}}{\sqrt{e^{2(d-1)A}-e^{2(d-1)A(r_*)}}}\ .
\end{equation}
We can show that the mutual information grows as the volume when the width of the strip is large enough. This happens when the disconnected surfaces approach each other at the wormhole's throat, $r_*\to 0$. Let us expand the warp factor around the minimum
\begin{equation}
    A(r)\sim A(0)+\frac{1}{2}A''(0) r^2\ .
\end{equation}
The leading contributions to the integrals that determine the mutual information and width are
\begin{equation}
\begin{split}
    2 (S_d(\text{A})-S_c(\text{A})) &\sim \frac{V(T^{d-2})}{G_{d+1}}e^{(d-2)A(0)}\int_{r_*}^{r_\epsilon} \dd r\, \frac{1}{\sqrt{(d-1)A''(0)(r^2-r_*^2)}}
    \ ,\\
    \Delta\ell(\text{A}) &\sim 2e^{-A(0)}\int_{r_*}^{r_\epsilon} \dd r\, \frac{1}{\sqrt{(d-1)A''(0)(r^2-r_*^2)}}\ ,
\end{split}
\end{equation}
where we have taken $r_\epsilon>r_*$ fixed and such that $A''(0)r_\epsilon^2\ll 1$. The integrals can be done analytically, and for $r_*\to 0$ they diverge as
\begin{equation}
\begin{split}
    2 (S_d(\text{A})-S_c(\text{A})) &\sim \frac{V(T^{d-2})}{G_{d+1}}\frac{e^{(d-2)A(0)}}{\sqrt{(d-1)A''(0)}}\log\frac{r_\epsilon}{r_*}
    \ ,\\
    \Delta\ell(\text{A}) &\sim 2\frac{e^{-A(0)}}{\sqrt{(d-1)A''(0)}}\log\frac{r_\epsilon}{r_*}\ .
\end{split}
\end{equation}
Thus, when the width is very large, the mutual information grows with the volume as
\begin{equation}\label{eq:mutualinfolarge}
    I(\text{A})\sim \frac{e^{(d-1)A(0)}}{2G_{d+1}}V(T^{d-2})\Delta \ell(\text{A})\ .
\end{equation}
Note that, in this limit, the factor $e^{(d-1)A(0)}V(T^{d-2})\Delta \ell(\text{A})$ becomes equal to the area of the wormhole throat at the point where the warp factor is minimal. Therefore, the amount of entanglement, as measured by the mutual information per unit volume, is determined by the `radius' of the wormhole throat.  This corresponds to the saturation of the bound relating the mutual information to the entanglement wedge cross section $E_W$ \cite{Freedman:2016zud,Takayanagi:2017knl}
\begin{equation}
    E_{W}(\rho_{\text{AB}}) \geq \frac{1}{2}I(\text{A}:\text{B})\ .
\end{equation}
In our case, $\text{A}$ and $\text{B}$ correspond to the same strip on each boundary of the wormhole and $\rho_{\text{AB}}$ denotes the corresponding reduced density matrix. The holographic entanglement wedge \cite{Czech:2012bh,Wall:2012uf,Headrick:2014cta} is the region delimited by the strip on the two boundaries and the minimal surfaces that determine the entanglement entropy of $\text{A}\cup\text{B}$. The entanglement wedge cross section is defined as the minimal area that divides the entanglement wedge into two parts, each containing the strip on one of the boundaries.

%%%%%%%%%%%%%%%%%%%%%%%%%%%%%%%%%%%%%%%%%%%%%%%%%%%%%%%%%
\subsubsection{Examples}\label{subsec:examples2}
%%%%%%%%%%%%%%%%%%%%%%%%%%%%%%%%%%%%%%%%%%%%%%%%%%%%%%%%%

\begin{figure}[t!]
    \centering
    \includegraphics[width=0.7\linewidth]{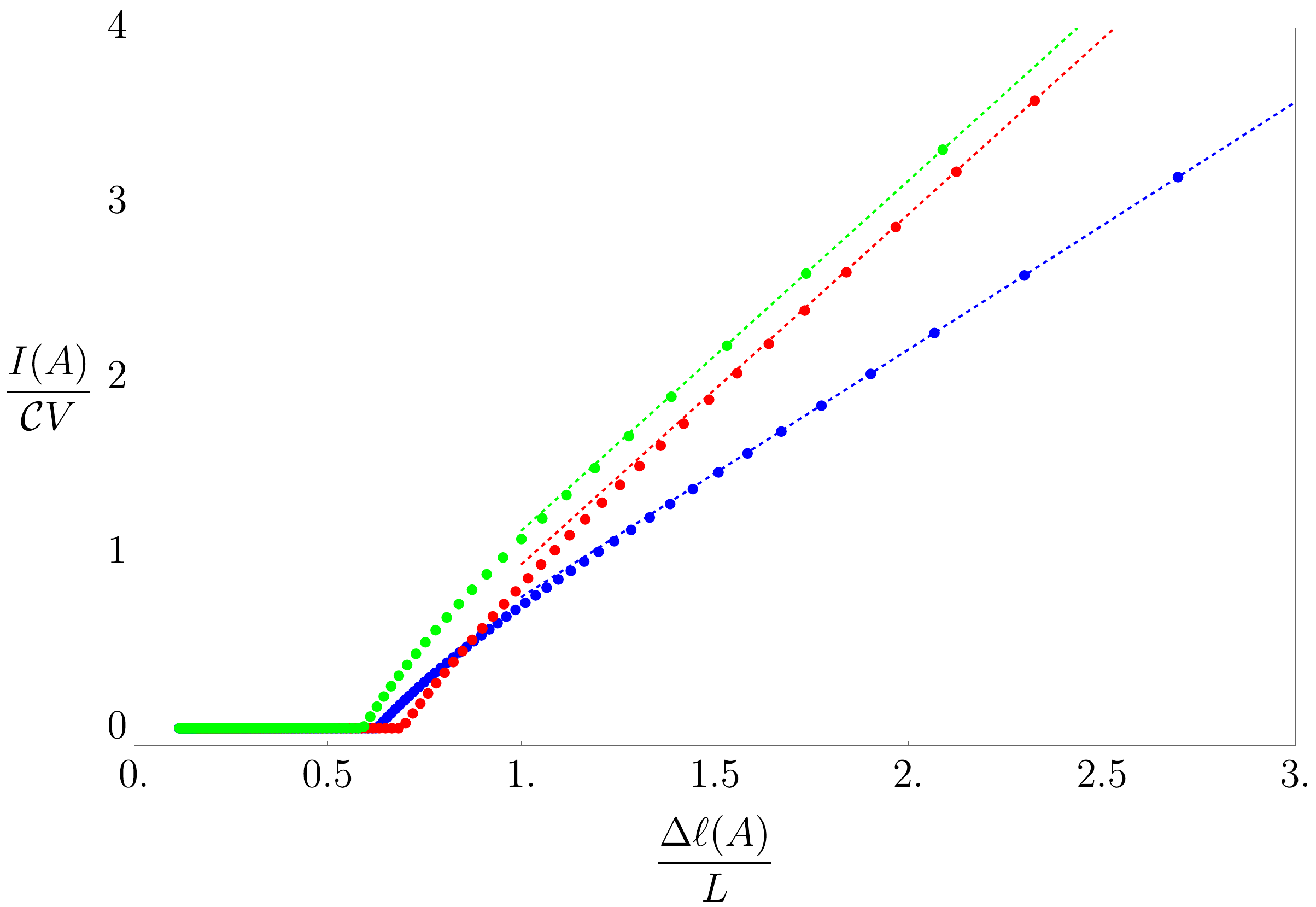}
    \caption{Mutual information of a strip $I(A)/({\cal C} V(T^{d-2}))$ as a function of the width $\Delta \ell(A)/L$, with ${\cal C}=L/G_{d+1}$. The cases represented here are the wormholes obtained from the analytic extensions of: 1. Coulomb branch of ${\cal N}=4$ SYM (blue), 2. Coulomb branch of ABJM (red), and 5. GPPZ (green). The dots are computed numerically using \eqref{eq:mutualinfstrip} and \eqref{eq:widthstrip}, and the dashed lines are linear fits at large values of the width. The slopes of the fits agree with \eqref{eq:mutualinfolarge}.}
    \label{fig:mutualinfo}
\end{figure}

For the wormhole solutions (Examples 1, 2 and 5 listed in Subsection \ref{subsec:examples}), we fix the boundary metric to be Minkowski at each asymptotic AdS region. The warp factor for each case is the following:
\begin{enumerate}
\item 
Asymptotic Minkowski corresponds to $r_0=0$ and $e^{2a_0}=2$ in \eqref{eq:sol1}. The warp factor reads
\begin{equation}
    e^{2A(r)}=2 \cosh\left({\frac{2r}{L}}\right)\ .
\end{equation}

\item 
Similarly, asymptotic Minkowski fixes $r_0=0$ and $e^{2a_0}=4$ in \eqref{eq:sol2} and the warp factor is
\begin{equation}
    e^{2A(r)}=4\cosh^2\left(\frac{r}{L}\right)\ .
\end{equation}

\item[5.] 
Here the asymptotic Minkowski condition fixes $r_1=r_2=0$ and $e^{2a_0}=2^{4/3}$ in \eqref{eq:sol5}. The warp factor is
\begin{equation}
    e^{2A(r)}=2^{4/3}\cosh\left(\frac{r}{L}\right)\left[\cosh\left(\frac{3r}{L}\right)\right]^{1/3}\ .
\end{equation}

\end{enumerate}
In the three examples above, the integrals \eqref{eq:mutualinfstrip} and \eqref{eq:widthstrip} can be computed numerically upon changing the radial variable to $u=r/L$. Results are plotted in Figure~\ref{fig:mutualinfo}.
In all three cases, the mutual information presents the following features: it vanishes below a (case-dependent) critical width, and it grows linearly with the width at large values. Fitting the high $\Delta\ell$ part of the plot to a linear function, $\frac{I}{\cal{C}V}=m\cdot \frac{\Delta \ell}{L} + n$, we find for each example:
\begin{align}
    m_1 &\approx 1.41427\,, \hspace{2.5cm} n_1 \approx -0.66611\,, \\
    m_2 &\approx 2.00003\,, \hspace{2.5cm} n_2 \approx -1.06581\,, \\
    m_5 &\approx 2.00007\,, \hspace{2.5cm} n_5 \approx -0.87333\,.
\end{align}
These values for the asymptotic slopes at large width agree with the formula in \eqref{eq:mutualinfolarge}.

%%%%%%%%%%%%%%%%%%%%%%%%%%%%%%%%%%%%%%%%%%%%%%%%%%%%%%%%%%%%%
%%%%%%%%%%%%%%%%%%%%%%%%%%%%%%%%%%%%%%%%%%%%%%%%%%%%%%%%%%%%%
\subsection{Spontaneous breaking of Poincar\'e symmetry to the diagonal group}
%%%%%%%%%%%%%%%%%%%%%%%%%%%%%%%%%%%%%%%%%%%%%%%%%%%%%%%%%%%%%
%%%%%%%%%%%%%%%%%%%%%%%%%%%%%%%%%%%%%%%%%%%%%%%%%%%%%%%%%%%%%

We have argued that the wormholes describe the low-energy limit of two stacks of branes and anti-branes that have undergone tachyon condensation. The endpoint consists of two decoupled copies of a QFT, but the wormhole is dual to a non-trivial entangled state in the product Hilbert space. 

In principle, two decoupled QFTs have two independent global Poincar\'e symmetries transforming the fields in each theory independently. We claim that the wormhole connecting the two asymptotic AdS boundaries is dual to a state that spontaneously breaks the symmetry to a diagonal symmetry $ISO(d-1,1)_1\times ISO(d-1,1)_2\to ISO(d-1,1)_D$.  The diagonal Poincar\'e symmetry is manifest in the isometries of the wormhole solution, and we show that there is a Goldstone boson corresponding to the broken symmetry.  Similar arguments have been made for Euclidean wormholes in \cite{Betzios:2023obs}, building on the solutions of \cite{Betzios:2019rds,Betzios:2021fnm}.

As we already mentioned, the non-gravitational analog of a wormhole provided by the connected D8--anti-D8-brane configuration in the Sakai--Sugimoto model corresponds to the spontaneous breaking of left and right chiral flavor symmetries to the diagonal vector symmetry $U(N_f)_L\times U(N_f)_R\to U(N_f)_V$, where $N_f$ is the number of flavors. Let us review in more detail how the holographic dictionary works in that example, as we employ it to inform our interpretation of the gravitational wormholes.

We denote by $z$ the radial holographic coordinate along the D8-brane worldvolume, such that the asymptotic boundaries are at $z\to \pm\infty$. On the connected D8-branes, there is a $U(N_f)$ gauge field. Sources for the global currents associated with each of the chiral flavor symmetries, $l_\mu$ and $r_\mu$, map, in the holographic dual, to the asymptotic values of the gauge field:
\begin{equation}
    \lim_{z\to-\infty} A_\mu(z,x)=l_\mu(x)
    \ ,\qquad 
    \lim_{z\to+\infty} A_\mu(z,x)=r_\mu(x)\ .
\end{equation}
Similarly, the expectation values of the chiral flavor currents can be obtained from the coefficients of the subleading terms in the asymptotic expansions of $A_\mu$ at each of the boundaries. The solutions for the gauge field split according to their parity in the $z$ direction. Parity-even solutions correspond to the vector sector, while parity-odd solutions correspond to the axial sector, the former being symmetric with respect to swapping left and right chiralities while the latter is antisymmetric. 

Proper gauge transformations in the D8-brane worldvolume are those vanishing at the boundaries. On the other hand, while gauge transformations that asymptote to a constant correspond to global symmetry transformations in the dual field theory, general gauge transformations that do not vanish at the boundary correspond to background gauge field transformations of the sources. 

Let us focus on the case where the gauge field vanishes in the background D8-brane solution and we look at linearized fluctuations around this background. The linearized gauge transformations are
\begin{equation}
    \delta A_M=\partial_M \lambda\ .
\end{equation}
We can introduce a Wilson line, invariant under proper gauge transformations, that connects the two boundaries 
\begin{equation}
     W(x)=\int_{-\infty}^\infty \dd z\, A_z(z,x)\ .
\end{equation}
For general gauge transformations, the Wilson line transforms as expected for a Goldstone boson of the broken axial symmetry
\begin{equation}
    \delta W(x)=\lim_{z\to +\infty} \lambda-\lim_{z\to-\infty} \lambda=\lambda_R-\lambda_L
    \ .
\end{equation}
By solving the equations of motion for fluctuations on the brane, one finds that indeed there is a massless mode making the Wilson line non-zero.

Let us now move on to the gravitational wormhole and show that there is a Goldstone associated with the broken Poincar\'e symmetry analogous to the Goldstone of the flavor symmetries of the D8-branes in the Sakai--Sugimoto model. The holographic radial coordinate is now denoted as $r$, with the asymptotic AdS boundaries at $r\to \pm\infty$.

In this case, we are interested in the metric. We could couple each of the dual QFTs to a different background metric, which would map to the asymptotic value of the metric in the gravity dual at the boundaries,
\begin{equation}
    \lim_{r\to-\infty} g_{\mu\nu}(r,x)\sim e^{-2r/L}g^{(0),[1]}_{\mu\nu}(x)
    \ ,\qquad
    \lim_{r\to+\infty} g_{\mu\nu}(r,x)\sim e^{2r/L}g^{(0),[2]}_{\mu\nu}(x)\ .
\end{equation}
The subleading terms in the asymptotic expansion at each boundary determine the expectation values of the energy momentum tensors in each of the QFTs. Following the example of the Sakai--Sugimoto model, if we classify solutions according to their parity with respect to the $r\to -r$ transformation, parity-even solutions should correspond to the sector of operators arranged in representations of the diagonal Poincar\'e subgroup and parity-odd solutions to the sector corresponding to the antidiagonal subgroup. 

The gauge symmetries associated with the metric are diffeomorphisms. 
We describe the background wormhole metric in domain wall coordinates
\begin{equation}
    \dd s^2
    =
    \dd r^2+e^{2A(r)}\eta_{\mu\nu}\,\dd x^\mu \dd x^\nu\ ,
\end{equation}
and expand the metric in background plus linear fluctuations
\begin{equation}
    G_{MN}=g_{MN}+h_{MN}\ .
\end{equation}
In the following, we do not enforce any gauge condition. Imposing the radial gauge would eliminate the components that capture the Goldstone mode.

Linearized diffeomorphisms transform the metric components as (indices are lowered with the Minkowski metric)
\begin{eqnarray}
    \delta h_{\mu\nu}&=&-2e^{2A}\partial_{(\mu}\xi_{\nu)}-2A'e^{2A}\eta_{\mu\nu}\xi^r\ ,\\
    \delta h_{\mu r} &=& -e^{2A}\partial_r \xi_\mu-\partial_\mu \xi^r\ ,\\
    \delta h_{rr} &=& -2\partial_r\xi^r\ .
\end{eqnarray}
We study vector fluctuations of the metric components $h_{\mu r}$ around the background. The other components are invariant under a subset of the diffeomorphisms satisfying
\begin{equation}\label{eq:alloweddiff}
     \xi^r=0
     \ ,\qquad \partial_{(\mu}\xi_{\nu)}=0
     \ .
\end{equation}
Thus, for each fixed value of $r$, $\xi^\mu(r,x)$ is a Killing vector of the Minkowski metric, which includes the generators of spacetime translations and Lorentz transformations,
\begin{equation}
    \xi_\mu=c_\mu+\omega_{\mu\nu} x^\nu
    \ ,\qquad
    \omega_{\mu\nu}=-\omega_{\nu\mu}\ .
\end{equation}
We consider the transverse vector component of the metric 
\begin{equation}
    H^\mu=-g^{\mu\alpha}h_{\alpha r}^t
    \ ,\qquad 
    \partial^\mu h_{\mu r}^t=0\ .
\end{equation}
Under linearized diffeomorphisms satisfying \eqref{eq:alloweddiff}, it transforms as
\begin{equation}
    \delta H^\mu=\partial_r \xi^{t\,\mu}
    \ ,\qquad
    \partial_\mu \xi^{t\,\mu}=0\ . 
\end{equation}
Note that the allowed diffeomorphisms \eqref{eq:alloweddiff} automatically satisfy the transverse condition, so we will drop the label below.

The analog to the Wilson line for translations is
\begin{equation}\label{eq:Wm}
    W^\mu(x)=\int_{-\infty}^\infty \dd r\, H^\mu(r,x)\ .
\end{equation}
Its transformation under diffeomorphisms is 
\begin{equation}
    \delta W^\mu=\lim_{r\to +\infty} \xi^\mu -\lim_{r\to -\infty} \xi^\mu\ .
\end{equation}
This is consistent with $W^\mu$ transforming as the Goldstone boson of the broken antidiagonal Poincar\'e symmetry, in particular translations. For Lorentz transformations we can introduce
\begin{equation}
    W_{\alpha\beta}(x)=2\partial_{[\alpha} W_{\beta]}(x)\ ,
\end{equation}
which transforms as expected for the Goldstones of the broken antidiagonal Lorentz symmetry
\begin{equation}
    \delta W_{\alpha\beta}=\lim_{r\to +\infty} 2\partial_{[\alpha}\xi_{\beta]} -\lim_{r\to -\infty} 2\partial_{[\alpha}\xi_{\beta]}\ .
\end{equation}
As usual, these are not independent from the Goldstones of broken translations \cite{Low:2001bw}.

Setting $h_{rr}=0$ and $h_{\mu\nu}=0$,
the linearized equations of motion for the vector metric fluctuation are
\begin{subequations}
\begin{eqnarray}
     \partial^2 h_{\mu r} &= &0\ ,\\
    \partial^\mu h_{\mu r} &= &0\ ,\\
     \partial_r \left(e^{(d-2)A} \partial_{(\mu} h_{\nu) r}\right) &=&0\ .
\end{eqnarray}
\end{subequations}
The first two equations correspond to a massless transverse vector mode along the Minkowski directions. From the last equation we obtain the radial dependence
\begin{equation}
    h_{\mu r}(r,x)=v_\mu(x) e^{-(d-2) A(r)}\ .
\end{equation}
This is a normalizable mode at both boundaries, in such a way that \eqref{eq:Wm} is finite. We can then identify this mode with the Goldstone for the broken antidiagonal Poincar\'e symmetry.

%%%%%%%%%%%%%%%%%%%%%%%%%%%%%%%%%%%%%%%%%%%%%%%%%%%%%%%%%
%%%%%%%%%%%%%%%%%%%%%%%%%%%%%%%%%%%%%%%%%%%%%%%%%%%%%%%%%
\section{Discussion}\label{sec:discuss}
%%%%%%%%%%%%%%%%%%%%%%%%%%%%%%%%%%%%%%%%%%%%%%%%%%%%%%%%%
%%%%%%%%%%%%%%%%%%%%%%%%%%%%%%%%%%%%%%%%%%%%%%%%%%%%%%%%%

Pseudo-Hermitian theories provide an alternative framework to define consistent quantum mechanical systems. It might be possible and very interesting to find pseudo-Hermitian formulations of quantum gravity by taking advantage of the holographic duality. We have taken a step in this direction by constructing solutions that can be interpreted as dual to pseudo-Hermitian theories in analytic extensions of well-known consistent supergravity truncations. The fact that we have found that some of the wormhole solutions, with a single ghost field, are linearly stable is encouraging, although we cannot rule out non-perturbative instabilities, or instabilities in fields not belonging to the consistent truncations. On the other hand, in an example with two coupled ghosts, we found an unstable mode.

In \cite{Maldacena:2026jqd} (see also \cite{DiUbaldo:2026rly,Etheredge:2026rio}) it has been argued that the distance along the imaginary direction that the scalar travels when traversing a wormhole places a bound on the analytic extension of a coupling in the dual field theory. The lower and upper bounds that were found for asymptotically $\text{AdS}_{d+1}$ boundaries were
\begin{equation}
    \text{IDB}_{\text{lower}}(d)
    =
    \pi \sqrt{\frac{2(d-1)}{d}}
    \ , \qquad 
    \text{IDB}_{\text{upper}}(d)
    =
    \pi \sqrt{\frac{2d}{d-1}}\ .
\end{equation}
These bounds were found for massless scalar fields, which are dual to marginal operators. In the examples we have constructed, the scalar operators have a non-trivial potential and the dual operators are relevant. We find that in these cases the distance between the two sides of the wormhole is larger than the upper bounds of the marginal case
\begin{align}
\text{Example 1:}\quad& \Delta \eta=\sqrt{3}\pi> \text{IDB}_{\text{upper}}(4)=2\sqrt{\frac{2}{3}}\pi
\ ,\\
\text{Example 2:}\quad& \Delta \eta=2\pi> \text{IDB}_{\text{upper}}(3)=\sqrt{3}\pi
\ ,\\
\text{Example 5:}\quad& |\Delta \vec \eta|= \sqrt{(\sqrt{3}\pi)^2+\pi^2} = 2\pi > \text{IDB}_{\text{upper}}(4)=2\sqrt{\frac{2}{3}}\pi\ .
\end{align}

Fluctuations of the wormhole solutions split according to their parity under reflection along the holographic direction, which exchanges the two boundaries. In the dual theory, parity-even and parity-odd modes would be naturally interpreted as diagonal and antidiagonal combinations of the two copies of the energy-momentum tensor and scalar operators.

The wormhole solutions we have found are Lorentzian, traversable, with boundaries that have flat metrics. In our interpretation of the solutions, the two boundaries do not correspond to two disconnected spacetimes, but to two copies of a theory in the same spacetime. The wormhole corresponds to an entangled state of the product Hilbert space. The state spontaneously breaks the two copies of Poincar\'e symmetry to their diagonal subgroup. If the origin of the breaking is tachyon condensation, the diffeomorphism Wilson line operator connecting the two boundaries would correspond to the components of the tachyon along flat directions, acting as the Goldstone bosons of the broken Poincar\'e symmetry. Extending the argument to other symmetries could serve as an alternative resolution---at the level of effective theory---of the factorization problem addressed in \cite{Harlow:2015lma,Guica:2015zpf}. Interestingly, in addition to the natural extension to Euclidean wormholes \cite{Betzios:2023obs}, spontaneous symmetry breaking due to entanglement between two boundaries of a holographic spacetime may be realized more generally. A similar mechanism seems to be at play in holographic entanglement island constructions, where the would-be Goldstone is eaten by a dynamical gauge field at one of the boundaries \cite{Geng:2025byh,Geng:2025gqu,Geng:2025gns,Geng:2025rov}.

Although we have focused on wormhole solutions, some of the other solutions that we have found have interesting characteristics, too. In particular, the flows to asymptotically flat spacetime have an uplift to ten-dimensional singular solutions with a real metric and imaginary fluxes. Example 6 of analytic extension of a supergravity solution, dual to a $3+1$-dimensional QFT, is uplifted to a type IIB solution with purely imaginary two-form NS-NS and R-R potentials. It is interesting to compare with the classification of \cite{Bergshoeff:2007cg}, which is summarized in their table 1. In addition to type IIB, they identify two other theories with real action, IIB${}^*$ and $\text{IIB}^\prime$ (related to each other through S-duality). In both cases, one of the two-form potentials is real and the other is imaginary, so the solution we have found does not belong to either. It might be that there is a more general classification if one relaxes some of the conditions, such as maximal supersymmetry. Example 7 is similar. It is the analytic extension of a supergravity solution dual to a $2+1$-dimensional QFT. The uplift is to an eleven-dimensional supergravity solution with four-form potential that has both real and purely imaginary components. This falls outside the classification of M-theory solutions given in Appendix D of \cite{Bergshoeff:2007cg}, where all the components would be either real or imaginary.

An interesting extension of this work would be to construct solutions interpolating between anti-de Sitter and de Sitter spacetimes. This could be achieved if there are flows that interpolate between critical points of the superpotential with negative and positive values. Recent works show that this is not possible quite generically if the null energy condition is satisfied \cite{Kiritsis:2019wyk,Kiritsis:2025ytb}. Since the analytic extension of the scalars introduces ghosts that violate the null energy condition, it is possible that this obstruction can be circumvented.

%%%%%%%%%%%%%%%%%%%%%%%%%%%%%%%%%%%%%%%%%%%%%%%%%%%%%%%%%
\section*{Acknowledgements}
%%%%%%%%%%%%%%%%%%%%%%%%%%%%%%%%%%%%%%%%%%%%%%%%%%%%%%%%%
%
We thank Daniel Areán, Alessio Caddeo, Adolfo Guarino and Pablo Soler for many useful discussions.  We also thank Panos Betzios, Hao Geng, Niko Jokela and Olga Papadoulaki for useful comments. This work is partially supported by the Spanish Agencia Estatal de Investigación and Ministerio de Ciencia, Innovación y Universidades through the grants PID2021-123021NB-I00 and PID2024-161500NB-I00. \\
The issues in the calculation of the spectrum were uncovered when Claude Opus 5.5 High was used to do a numerical calculation in a related project. We created the Mathematica code used to compute the modes in the three examples, with independent numerical verifications of Examples 1 and 5 also done by Claude. The authors have checked the code and the text and take full responsibility on the contents and results.

\appendix

%%%%%%%%%%%%%%%%%%%%%%%%%%%%%%%%%%%%%%%%%%%%%%%%%%%%%%%%%
%%%%%%%%%%%%%%%%%%%%%%%%%%%%%%%%%%%%%%%%%%%%%%%%%%%%%%%%%
\section{A general type IIB truncation}\label{app:trunc}
%%%%%%%%%%%%%%%%%%%%%%%%%%%%%%%%%%%%%%%%%%%%%%%%%%%%%%%%%
%%%%%%%%%%%%%%%%%%%%%%%%%%%%%%%%%%%%%%%%%%%%%%%%%%%%%%%%%

In this appendix we consider the truncation to six scalars that appeared in \cite{Khavaev:2000gb,Bobev:2010de}. Truncating different combinations of the scalars, one recovers most of the examples discussed in the main sections: the $\mathcal{N}=2^*$, Leigh--Strassler, GPPZ and holographic superconductor cases (numbered 3 to 6 above). Furthermore, it contains the vectors and phases that allow us to interpret four of them as the modulus of a complex, charged scalar, so that our general observations apply. The action can be written as 
\begin{align}
S=\frac{1}{2\kappa^2}\int\dd^5x & \sqrt{-g}\left[R-2\sum_{i=1}^4\partial_\mu\varphi_i\partial^\mu\varphi_i-12\partial_\mu\alpha\partial^\mu\alpha-4\partial_\mu\beta\partial^\mu\beta-\frac12\sum_{i=1}^4\sinh^2(2\varphi_i)D_\mu\theta_i D^\mu\theta_i\right.\nonumber\\[2mm]
&\left.-e^{4(\alpha-\beta)}F_{\mu\nu}^1F^{1\,\mu\nu}-e^{4(\alpha+\beta)}F_{\mu\nu}^2F^{2\,\mu\nu}-e^{-8\alpha}F_{\mu\nu}^3F^{3\,\mu\nu}-V\right]+S_{\rm top}\ ,
\end{align}
where $S_{\rm top}$ are topological terms for the vectors, not specified in \cite{Bobev:2010de}, that play no role in the following. We have defined the covariant derivatives
\begin{align}
D_\mu\theta_1=\partial_\mu\theta_1+A_\mu^1+A_\mu^2-A_\mu^3\ ,\qquad D_\mu\theta_2=\partial_\mu\theta_2+A_\mu^1-A_\mu^2+A_\mu^3\ ,\nonumber\\[2mm]
D_\mu\theta_3=\partial_\mu\theta_3-A_\mu^1+A_\mu^2+A_\mu^3\ ,\qquad D_\mu\theta_4=\partial_\mu\theta_4-A_\mu^1-A_\mu^2-A_\mu^3\ .
\end{align} 
The potential is given in terms of the superpotential\footnote{The constant $g$ appearing in \cite{Khavaev:2000gb,Bobev:2010de} is related to the AdS radius as $g=2/L$.} 
\begin{align}
W=-\frac{e^{-2\alpha-2\beta}}{L}&\left[\left(1+e^{4\beta}-e^{6\alpha+2\beta}\right)\cosh(2\varphi_1)+\left(-1+e^{4\beta}+e^{6\alpha+2\beta}\right)\cosh(2\varphi_2)\right.\nonumber\\[2mm]
&\left.+\left(1-e^{4\beta}+e^{6\alpha+2\beta}\right)\cosh(2\varphi_3)+\left(1+e^{4\beta}+e^{6\alpha+2\beta}\right)\cosh(2\varphi_4)\right]\ ,
\end{align}
as
\begin{align}
V&=\frac18\left[\sum_{i=1}^4\left(\frac{\partial W}{\partial\varphi_i}\right)^2+\frac16\left(\frac{\partial W}{\partial\alpha}\right)^2+\frac12\left(\frac{\partial W}{\partial\beta}\right)^2\right]-\frac13W^2\ .\nonumber
\end{align}
From the form of the action it is evident that the $\varphi_i$ scalars can be combined with the phases $\theta_i$ to form four charged, complex scalars $\zeta_j=\tanh{(\varphi_j)}\,e^{\im\theta_j}$, $j=1,\dots,4$, with kinetic terms 
\begin{align}
2\sum_{i=1}^4\cosh^4(\varphi_i)D_\mu\zeta_i\overline{D^\mu\zeta_i}\ .
\end{align}
The neutral scalars are dual to boson bilinear operators and thus have dimension $\Delta=2$. In terms of the usual six scalars $X_a$, $a=1,\dots,6$, of $\mathcal{N}=4$ super Yang--Mills, the relation is
\begin{align}
\alpha\quad&\leftrightarrow\quad \Tr\left(2X_6^2+2X_5^2-X_4^2-X_3^2-X_2^2-X_1^2\right),\nonumber\\[2mm]
\beta\quad&\leftrightarrow\quad\Tr\left(X_1^2+X_2^2-X_3^2-X_4^2\right).
\end{align}
On the other hand, the charged scalars are dual to fermion bilinears with dimension $\Delta=3$. Their moduli are identified as 
\begin{align}
\varphi_j\quad\leftrightarrow\quad\Tr\left(\lambda_j\lambda_j\right)+{\rm h.c.}\ ,
\end{align}
with $\lambda_j$ the four fermions of the gauge theory. 

This model has several interesting subtruncations that yield some of the models and solutions discussed in Subsections \ref{subsec:superpotential} and \ref{subsec:examples}. First, it is consistent to truncate away $\beta=\varphi_1=\varphi_4=0$ while identifying $\varphi_3=\varphi_2=\chi$.\footnote{This has to be accompanied with $\theta_1=\theta_4=0$, together with $A^1_\mu=A^2_\mu=0$.} The resulting model gives the $\mathcal{N}=2^*$ flow of \cite{Pilch:2000ue}. Indeed, the superpotential reduces to 
\begin{align}
W=-\frac{2}{L}\left[2e^{-2\alpha}+e^{4\alpha}\cosh(2\chi)\right],
\end{align}
which coincides with that of Example 3 above.

The Leigh--Strassler flow in \cite{Pilch:2000fu} can be recovered by truncating out $\varphi_2=\varphi_3=\varphi_4=\beta=0$ and keeping just $\alpha$ and $\varphi_1=\chi$. The superpotential reduces to 
\begin{align}
  W=-\frac{e^{-2\alpha}}{L}\left[\left(2-e^{6\alpha}\right)\cosh(2\chi)+2+3e^{6\alpha}\right],  
\end{align}
which is indeed that of Example 4. The resulting equations cannot be integrated in closed form and must be solved numerically. 

The GPPZ model \cite{Girardello:1999bd} can be obtained by fixing $\alpha=\beta=0$ and identifying the three scalars $\varphi_1=\varphi_2=\varphi_3=\frac{m}{\sqrt{3}}$, together with $\varphi_4=\sigma$. In addition, it is necessary to identify the phases, $\theta_1=\theta_2=\theta_3$, and the vectors, $A_\mu^1=A_\mu^2=A_\mu^3$. The superpotential reduces to
\begin{align}
W=-\frac{3}{L}\left[\cosh\left(\frac{2m}{\sqrt{3}}\right)+\cosh(2\sigma)\right],
\end{align}
giving case 5 above. It is of course possible to further truncate this model by setting either $m$ or $\sigma$ to zero. The first option gives rise to the model with a single charged scalar in \cite{Gubser:2009qm}, which corresponds to our Example 6.\footnote{It is possible to enrich this model keeping $\alpha$ in addition.}

Finally, it is of course consistent to switch off the complete set of charged scalars and keep the neutral ones $\alpha$ and $\beta$. The resulting model is our Example 8 above.

%%%%%%%%%%%%%%%%%%%%%%%%%%%%%%%%%%%%%%%%%%%%%%%%%%%%%%%%%
%%%%%%%%%%%%%%%%%%%%%%%%%%%%%%%%%%%%%%%%%%%%%%%%%%%%%%%%%
\section{Equations of motion}\label{app:eoms}
%%%%%%%%%%%%%%%%%%%%%%%%%%%%%%%%%%%%%%%%%%%%%%%%%%%%%%%%%
%%%%%%%%%%%%%%%%%%%%%%%%%%%%%%%%%%%%%%%%%%%%%%%%%%%%%%%%%

The action in the gravity dual consists of Einstein gravity plus ordinary and ghost scalars
\begin{equation}
    S_{d+1}=\frac{1}{16\pi G_{d+1}}\int \dd^{d+1} x\, \sqrt{-G}\left[R+\frac{1}{2}\partial_M \eta^a\partial^M\eta^a-\frac{1}{2}\partial_M \phi^i\partial^M\phi^i-V\left(\{\eta^a,\phi^i\}\right)\right]\ .
\end{equation}
Einstein's equations are
\begin{equation}
    R_{MN}=\frac{1}{2}\widetilde T_{MN}\equiv \frac{1}{2}\left(T_{MN}-\frac{1}{d-1}G_{MN}T^L_{\ L} \right)\ ,
\end{equation}
where
\begin{equation}
    T_{MN}=-\partial_M \eta^a\partial_N\eta^a+\partial_M \phi^i\partial_N\phi^i-\frac{1}{2}G_{MN}\left( -\partial_L \eta^a\partial^L\eta^a+\partial_L \phi^i\partial^L\phi^i+2V\right)\ ,
\end{equation}
hence
\begin{equation}
    \widetilde T_{MN}=-\partial_M \eta^a\partial_N\eta^a+\partial_M \phi^i\partial_N\phi^i+\frac{2}{d-1}G_{MN}V\ .
\end{equation}
The scalar equations are
\begin{equation}
    \nabla^M\partial_M\phi^i-\partial_{\phi^i}V=0
    \ ,\qquad \nabla^M\partial_M\eta^a+\partial_{\eta^a}V=0\ .
\end{equation}

%%%%%%%%%%%%%%%%%%%%%%%%%%%%%%%%%%%%%%%%%%%%%%%%%%%%%%%%%
\subsection{Equations for the background}
%%%%%%%%%%%%%%%%%%%%%%%%%%%%%%%%%%%%%%%%%%%%%%%%%%%%%%%%%

The background metric and scalars for a domain wall solution are 
\begin{equation}
    \dd s^2=g_{MN}\dd x^M \dd x^N=\dd r^2+e^{2A(r)}\eta_{\mu\nu}\dd x^\mu \dd x^\nu
    \ ,\qquad 
    \phi^i=\phi^i_0(r)
    \ ,\qquad 
    \eta^a=\eta_0^a(r)\ .
\end{equation}
Einstein's equations can be simplified to
\begin{equation}\label{eq:Einsbackg}
    A''+d(A')^2+\frac{V_0}{d-1}=0
    \ ,\qquad 
    (d-1)A''=\frac{1}{2}({\eta_0^a}')^2-\frac{1}{2}({\phi_0^i}')^2\ ,
\end{equation}
where $V_0=V(\{\eta^a_0,\phi^i_0\})$. The equations for the background scalars are
\begin{equation}\label{eq:scalarbackg}
    {\phi_0^i}''+dA'{\phi_0^i}'-\partial_{\phi^i}V_0=0
    \ ,\qquad {\eta_0^a}''+dA'{\eta_0^a}'+\partial_{\eta^a}V_0=0\ .
\end{equation}
One can check that solutions to the first order equations satisfy the second order equations as well
\begin{equation}
   A'=-\frac{W_0}{2(d-1)}
   \ ,\qquad {\phi^i}'=\partial_{\phi^i}W_0
   \ ,\qquad
   {\eta^a}'=-\partial_{\eta^a}W_0\ ,
\end{equation}
where $W_0=W(\{\eta^a_0,\phi^i_0\})$ and
\begin{equation}
    V=\frac{1}{2}(\partial_{\phi^i}W)^2-\frac{1}{2}(\partial_{\eta^a} W)^2-\frac{d}{4(d-1)}W^2\ .
\end{equation}

%%%%%%%%%%%%%%%%%%%%%%%%%%%%%%%%%%%%%%%%%%%%%%%%%%%%%%%%%
\subsection{Linearized fluctuations}
%%%%%%%%%%%%%%%%%%%%%%%%%%%%%%%%%%%%%%%%%%%%%%%%%%%%%%%%%

We will expand to linear order in the metric and the scalar fields around the domain wall background
\begin{equation}
    G_{MN}=g_{MN}+h_{MN}
    \ ,\qquad \phi^i=\phi^i_0+\varphi^i
    \ ,\qquad 
    \eta^a=\eta^a_0+\chi^a\ .
\end{equation}
The linear-order Ricci tensor has the form
\begin{equation}
   \delta R_{MN}= -\frac{1}{2}\left(\nabla^2 h_{MN}-2\nabla^L \nabla_{(M}h_{N)L}+\nabla_M \partial_N h\right)\ ,
\end{equation}
where covariant derivatives are with respect to the background metric, indices are raised and lowered with the background metric, and $h=g^{MN}h_{MN}$.\\
The energy-momentum tensor of the scalars to linear order is
\begin{equation}
\begin{split}
     \delta \widetilde T_{MN}=&-2\partial_{(M}\eta_0^a \partial_{N)}\chi^a+2\partial_{(M}\phi_0^i \partial_{N)}\varphi^i\\
     &+\frac{2}{d-1}h_{MN}V_0+\frac{2}{d-1}g_{MN}\partial_{\phi^i} V_0\,\varphi^i+\frac{2}{d-1}g_{MN}\partial_{\eta^a} V_0\,\chi^a\ .
    \end{split}
\end{equation}
Expanding to linear order the scalar equations of motion, we get
\begin{eqnarray}
\notag \nabla^M \partial_M \varphi^i-\partial_{\phi^i}\partial_{\phi^j} V_0\,\varphi^j-\partial_{\phi^i}\partial_{\eta^a} V_0\,\chi^a-h^{MN}\nabla_M\partial_N \phi_0^i-\left(\nabla_M h^{MN}-\frac{1}{2}\partial^N h\right)\partial_N\phi^i_0 &=&0, \\
& & \\
\notag \nabla^M \partial_M \chi^a+\partial_{\eta^a}\partial_{\eta^b} V_0\,\chi^b+\partial_{\eta^a}\partial_{\phi^i} V_0\,\varphi^i-h^{MN}\nabla_M\partial_N \eta_0^a-\left(\nabla_M h^{MN}-\frac{1}{2}\partial^N h\right)\partial_N\eta^a_0 &=&0.\\ & &
\end{eqnarray}
In the domain wall background the only non-zero Christoffel symbols are
\begin{equation}
    \Gamma^r_{\mu\nu}=-A' e^{2A}\eta_{\mu\nu}=-A' g_{\mu\nu}
    \ , \qquad 
    \Gamma^\mu_{r\nu}=A' \delta^\mu_\nu\ .
\end{equation}
Then, the scalar equations become, after using the background equations of motion
\begin{eqnarray}
\notag \partial^M \partial_M \varphi^i+dA'\partial_r\varphi^i -\partial_{\phi^i}\partial_{\phi^j} V_0\,\varphi^j-\partial_{\phi^i}\partial_{\eta^a} V_0\,\chi^a-\partial_{\phi^i} V_0 h_{rr} -\left(\partial^M h_{Mr}-\frac{1}{2}\partial_r h\right){\phi_0^i}' &=&0, \\ & & \\
\notag \partial^M \partial_M \chi^a+dA'\partial_r\chi^a +\partial_{\eta^a}\partial_{\eta^b} V_0\,\chi^b+\partial_{\eta^a}\partial_{\phi^i} V_0\,\varphi^i+\partial_{\eta^a} V_0 h_{rr} -\left(\partial^M h_{Mr}-\frac{1}{2}\partial_r h\right){\eta_0^a}' &=&0 .\\ \label{eq:chieq} & &
\end{eqnarray}
The components of the energy-momentum tensor are
\begin{eqnarray}
    \delta \widetilde T_{rr} &=& -2{\eta_0^a}'\partial_r \chi^a+2{\phi_0^i}'\partial_r \varphi^i+\frac{2}{d-1}\left(\partial_{\phi^i} V_0 \varphi^i+\partial_{\eta^a} V_0 \chi^a+ V_0 h_{rr}\right),\\
    \delta \widetilde T_{\mu r} &=& -{\eta_0^a}' \partial_\mu\chi^a+{\phi^i_0}'\partial_\mu \varphi^i+\frac{2}{d-1}V_0 h_{\mu r},\\
    \delta \widetilde T_{\mu\nu} &=& \frac{2}{d-1}g_{\mu\nu} \left(\partial_{\phi^i} V_0\, \varphi^i+\partial_{\eta^a} V_0\, \chi^a\right)+\frac{2}{d-1}V_0 h_{\mu\nu}.
\end{eqnarray}
The components of the Ricci tensor can be written as
\begin{eqnarray}
 \notag   -2\delta R_{rr} &=&\partial^M\partial_M h_{rr}-2\partial_r\left(\partial^M h_{Mr}-\frac{1}{2}\partial_rh\right)-4 A'\left(\partial^M h_{Mr} -\frac{1}{2}\partial_r h\right)\\
    & & -(d-2)A' \partial_r h_{rr}\ ,\\
 \notag    -2\delta R_{r\mu} &=& \partial^M\partial_M h_{r\mu}-\partial_r\left(\partial^M h_{M\mu}-\frac{1}{2}\partial_\mu h\right)-\partial_\mu\left(\partial^M h_{Mr}-\frac{1}{2}\partial_rh\right)+2 A'' h_{\mu r}
 \\
 & & + 2d(A')^2 h_{\mu r}-(d-1) A'\partial_\mu h_{rr}
 \ ,\\
\notag    -2\delta R_{\mu\nu} &=& \partial^M\partial_M h_{\mu\nu}-2\partial_{(\mu}\left(\partial^M  h_{\nu) M}-\frac{1}{2}\partial_{\nu)} h\right)-2 A'' g_{\mu\nu}h_{r r}\\
\notag & & +(A')^2\left(4 h_{\mu\nu}-2d g_{\mu\nu} h_{rr}\right)\\
 & & +A'\left[-2(d-2)\partial_{(\mu} h_{\nu)r}-2g_{\mu\nu} \left(\partial^M h_{Mr}-\frac{1}{2}\partial_r h\right)+(d-4)\partial_r h_{\mu\nu}\right]\ .
\end{eqnarray}
In the case of a metric vector fluctuation with $\varphi^i=0$, $\chi^a=0$, $h_{rr}=0$ and $h_{\mu\nu}=0$, the scalar equations and the Einstein equation for the $rr$ component become
\begin{equation}\label{eq:transvapp}
    \partial^\mu h_{\mu r}=0\ .
\end{equation}
The Einstein equations for the $r\mu$ components become
\begin{equation}
\begin{split}
0=& -2\delta R_{r\mu}+\delta \widetilde T_{r\mu}\\ &=\partial^M\partial_M h_{\mu r}-\partial_r^2 h_{r\mu}+2\left[A''+d(A')^2+\frac{1}{d-1} V_0\right] h_{r\mu}=g^{\alpha\beta}\partial_\alpha \partial_\beta h_{r\mu}\ .
\end{split}
\end{equation}
The equation for the $\mu\nu$ components is
\begin{align}
0=& -2\delta R_{\mu\nu}+\delta \widetilde T_{\mu\nu}
=-2\partial_r\left(\partial_{(\mu}h_{\nu)r}\right)-2(d-2)A'\partial_{(\mu}h_{\nu)r}-2A' g_{\mu\nu}\partial^\alpha h_{\alpha r}
\\ \nonumber&=-2e^{-(d-2)A}\partial_{(\mu}\partial_r\left( e^{(d-2)A} h_{\nu)r}\right),
\end{align}
where we used \eqref{eq:transvapp} in the last step.

A scalar fluctuation corresponds to the ansatz
\begin{equation}
 h_{rr}=h_R\ , \quad   h_{r\mu}=\partial_\mu h_V
    \ ,\qquad h_{\mu\nu}=e^{2A}\eta_{\mu\nu}h_S+\partial_\mu\partial_\nu h_L\ .
\end{equation}
A massless fluctuation would obey the equations
\begin{subequations}\label{eq:dilatonmode}
\begin{eqnarray}
    (r\mu): \ 0 &=& \partial_r h_S-A' h_R-\frac{\eta_0'}{d-1}\chi
    \ ,\\
    (rr): \ 0&=& \partial_r^2 h_S-A'\partial_r h_R-\frac{2}{d-1}\eta_0'\partial_r \chi
    \ ,\\
    \eta^{\mu\nu}(\mu\nu): \ 0&=&(d-2) A'' h_R+(\eta_0''\chi-\eta_0' \partial_r\chi)\ ,\\
    (\chi): \ 0&=& \partial_r^2 \chi+dA'\partial_r\chi+\partial_\eta^2 V\chi+\partial_\eta V h_R-\frac{1}{2}(\partial_r h_R-d\partial_r h_S)\eta_0'\ .
\end{eqnarray}
\end{subequations}
There is also an additional equation $\sim \partial_\mu\partial_\nu$ involving $h_V$ and $h_L$.
The difference of the first two equations gives
\begin{equation}
    (d-1)A''h_R+(\eta_0''\chi-\eta_0'\partial_r\chi)=0\ .
\end{equation}
Thus, we get
\begin{equation}\label{eq:dilatonsolution}
    h_R=0
    \ ,\qquad \chi=\tau(x) \eta_0'(r)
    \ ,\qquad
    \partial_r h_S=\frac{(\eta_0')^2}{d-1}\tau(x)\ .
\end{equation}
One can check that the last equation in \eqref{eq:dilatonmode} is satisfied using the first order equations and the definition of the potential in terms of the superpotential. Usually this would be the dilaton mode, but, if we integrate the equation for $h_S$ in \eqref{eq:dilatonsolution}, we cannot fix the integration constant in such a way that $h_S$ vanishes at both boundaries. Thus, the would-be dilaton mode cannot be made normalizable at both boundaries simultaneously.

%%%%%%%%%%%%%%%%%%%%%%%%%%%%%%%%%%%%%%%%%%%%%%%%%%%%%%%%%
\subsubsection{Equations in radial gauge for a single ghost field}
%%%%%%%%%%%%%%%%%%%%%%%%%%%%%%%%%%%%%%%%%%%%%%%%%%%%%%%%%

In order to study the stability of the solutions when there is a single ghost field we will work in the radial gauge $h_{M r}=0$ and expand in plane waves along the field theory directions
\begin{equation}
\begin{split}
    &h_{\mu\nu}(r,x)=\int \frac{\dd^dk}{(2\pi)^d} e^{ik\cdot x}H_{\mu\nu}(r,k),\\
    &\chi(r,x)=\int \frac{\dd^dk}{(2\pi)^d} e^{ik\cdot x}X(r,k).
\end{split}
\end{equation}
In the following we will use $\partial_r=\,'$ for the Fourier components, as we did for the background solutions. 

The equations of motion are invariant under linearized diffeomorphisms
\begin{equation}
    \delta h_{MN}=-g_{ML}\partial_N \xi^L-g_{NL}\partial_M \xi^L-\xi^L\partial_L g_{MN},\quad \delta \chi=-\xi^L\partial_L\eta_0.
\end{equation}
Fixing the radial gauge does not completely fix all diffeomorphisms; there is a remnant which takes the form
\begin{equation}
    \xi^r(r,x)=\xi^r(x),\quad \xi^\mu(r,x)=\eta^{\mu\nu}\left(\lambda_\nu(x)-\partial_\nu\xi^r(x)\int_{r_0}^r \dd u\, e^{-2A(u)}\right).
\end{equation}
The transformation of the fluctuations under the remnant diffeomorphism is
\begin{equation}
    \delta \chi=-\xi^r\eta_0'
    \ ,\quad 
    \delta h_{\mu\nu}=-2e^{2A}\partial_{(\mu}\lambda_{\nu)}-2 A' e^{2A}\eta_{\mu\nu}\xi^r+2\partial_\mu\partial_\nu\xi^r e^{2A}\int_{r_0}^r \dd u\, e^{-2A(u)}\ .
\end{equation}
We can construct invariant combinations by adding some of the fluctuations with appropriate factors. In order to simplify the analysis, let us work with the modes in the plane wave expansion, and introduce the following $d-1$ transverse polarization vectors with respect to the (timelike) momentum $k^\mu=(\omega,\bm{k})$,
\begin{equation}
q_\mu=\frac{1}{\sqrt{\omega^2-\bm{k}^2}}\left(|\bm{k}|\ ,-\omega\frac{\bm{k}}{|\bm{k}|}\right)\ ,\quad e_\mu^\alpha=(0,\bm{e}^\alpha), \ \alpha=1,\dots,d-2\ ,
\end{equation}
where
\begin{equation}
     \bm{k}\cdot\bm{e}^\alpha=0
     \ ,\qquad \bm{e}^\alpha\cdot\bm{e}^\beta=\delta^{\alpha\beta}\ .
\end{equation}
In the following, we will use the Minkowski metric to raise and lower indices along the field theory directions. There are two independent scalar invariant combinations
\begin{eqnarray}
 \label{eq:zetaeta}   
 Z_\eta=X-\frac{\eta_0'e^{-2A}}{2A'}\frac{1}{d-2}\sum_{\alpha=1}^{d-2}e_\mu^\alpha e_\nu^\alpha H^{\mu\nu}
 ,\\
 \label{eq:HS}
 H_S=\left(q_\mu q_\nu-\frac{1}{d-2}\sum_{\alpha=1}^{d-2}e_\mu^\alpha e_\nu^\alpha\right)H^{\mu\nu}.
\end{eqnarray}
There are $d-2$ invariant vector modes 
\begin{equation}
    H_V^\alpha=q_\mu e_\nu^\alpha H^{\mu\nu}.
\end{equation}
Finally, the invariant tensor modes are, for $\alpha< \beta$,
\begin{equation}
\begin{split}
&H_1^{\alpha\beta}=H_1^{\beta\alpha}=e^{\alpha}_\mu e^{\beta}_\nu H^{\mu\nu},\\
&H_2^{\alpha\beta}=-H_2^{\beta\alpha}=(e_\mu^{\alpha}e_\nu^{\alpha}-e_\mu^{\beta}e_\nu^{\beta})H^{\mu\nu}.
\end{split}
\end{equation}
This gives $\frac{(d-2)(d-1)}{2}-1$ independent tensor modes. In total there are $1+\frac{(d-2)(d+1)}{2}$
invariant modes, which agrees with the $1+\frac{d(d+1)}{2}$ fluctuations of the metric and scalar, minus the $d+1$ constraints that we obtain fixing the radial gauge. The constraints remove the $d$ longitudinal components of the metric and the trace part of the transverse components.

We can rearrange the equations of the linearized fluctuations to obtain a set of minimally coupled equations for the invariant modes in the radial gauge. The equations for the invariant modes not involving the scalar field are found simply by contracting the $\mu\nu$ Einstein equations with the projectors that appear in their definitions. Taking also into account that these modes are transverse and traceless, one gets that each of them is decoupled from the rest of the modes and satisfies the same equation ($H=H_S, H_V^\alpha,H_1^{\alpha\beta},H_2^{\alpha\beta}$)
\begin{equation}
    H''+(d-4)A' H'+\left[e^{-2A}M^2+4(A')^2+\frac{2}{d-1}V_0\right]H=0\ ,
\end{equation}
where we have defined $k^2=k^\mu k_\mu=-M^2$. We can combine \eqref{eq:Einsbackg} to replace the potential evaluated in the background solution by
\begin{equation}
    V_0=-d(d-1)(A')^2-\frac{1}{2}(\eta_0')^2\ .
\end{equation}
The final form of the equation is
\begin{equation}\label{eq:invariantHeq}
    H''+(d-4)A' H'+\left[e^{-2A}M^2-2(d-2)(A')^2-\frac{1}{d-1}(\eta_0')^2\right]H=0\ .
\end{equation}

To derive the equation for the remaining mode, we introduce
\begin{equation}
    \widehat H_S=\frac{1}{d-2}\sum_{\alpha=1}^{d-2}e_\mu^\alpha e_\nu^\alpha H^{\mu\nu}, \quad H_L=\frac{k_\mu k_\nu}{k^2}H^{\mu\nu},
\end{equation}
in such a way that the trace is $h=e^{-2A}(H_S+(d-1)\widehat H_S+H_L)$. Then, from the equation for the ghost scalar \eqref{eq:chieq}, the $\mu r$ Einstein equations projected along the momentum, and the $\mu\nu$ Einstein equations projected on $\widehat H_S$ we get the set of equations
\begin{eqnarray}
\label{eq:scalarX} 0&=& X''+dA'X'+e^{-2A}M^2 X+\partial_\eta^2 V_0 X+\frac{1}{2}\eta_0' h'=0,\\
\label{eq:scalarCons} 0&=& H_S'-2A' H_S+(d-1)(\widehat H_S'-2A' \widehat H_S)-e^{2A}\eta_0' X,\\
\label{eq:scalarhHS} 0&=& \widehat H_S''+(d-4) A' \widehat H_S'+\left[4(A')^2+\frac{V_0}{d-1}\right]\widehat H_S+\frac{2}{d-1}e^{2A}\partial_\eta V_0 X+e^{2A}A' h'.
\end{eqnarray}
We subtract \eqref{eq:scalarX}$-\frac{\eta_0'e^{-2A}}{2A'}\eqref{eq:scalarhHS}$ identifying the invariant combination \eqref{eq:zetaeta}. We remove additional terms proportional to $\widehat H_S'$ solving for it in \eqref{eq:scalarCons}. The remaining terms depend on $H_S'$, $H_S$ and $\widehat H_S$, but using the equations for the background \eqref{eq:Einsbackg}, \eqref{eq:scalarbackg} one can show that the coefficient of the $\widehat H_S$ term cancels out, so one is left with an equation depending only on invariant quantities
\begin{equation}\label{eq:invariantZeq}
    0=Z_\eta''+dA' Z_\eta'+\left(e^{-2A}M^2+Q_{ZZ}\right)Z_\eta+Q_{ZH}(H_S'-2A' H_S),
\end{equation}
with
\begin{eqnarray}\label{eq:Qcoefs}
    Q_{ZZ}&=& \partial_\eta^2V_0-\frac{1}{d-1}\frac{\eta_0'}{A'}\partial_\eta V_0-\eta_0' e^{2A} Q_{ZH},\\
    Q_{ZH} &=& \frac{e^{-2A}}{d-1}\frac{\eta_0'A''-A'\eta_0''}{(A')^2}.
\end{eqnarray}
Using the background equations of motion \eqref{eq:Einsbackg}, \eqref{eq:scalarbackg} one can also write
\begin{equation}\label{eq:Qcoef2}
    \eta_0'A''-A'\eta_0''=A'\partial_\eta V_0+d(A')^2\eta_0'+\frac{(\eta_0')^3}{2(d-1)}.
\end{equation}

%%%%%%%%%%%%%%%%%%%%%%%%%%%%%%%%%%%%%%%%%%%%%%%%%%%%%%%%%
%%%%%%%%%%%%%%%%%%%%%%%%%%%%%%%%%%%%%%%%%%%%%%%%%%%%%%%%%
\section{Spectrum of ghost scalar fluctuations}\label{app:numerics}
%%%%%%%%%%%%%%%%%%%%%%%%%%%%%%%%%%%%%%%%%%%%%%%%%%%%%%%%%
%%%%%%%%%%%%%%%%%%%%%%%%%%%%%%%%%%%%%%%%%%%%%%%%%%%%%%%%%

This appendix provides some details on the explicit computation of the low-lying values for the squared mass $M^2$ in the spectrum of normal modes for Examples 1, 2 and 5 listed in Subsection \ref{subsec:examples}.%
\footnote{
At \cite{code} we share a Wolfram Mathematica notebook collecting the actual computations described in this appendix.
}
The normalization of the warp factor is $e^{2a_0}=2,4, 2^{4/3}$ for Examples 1, 2 and 5, respectively, as specified in Subsection \ref{subsec:examples2}.\\

\noindent 
\textbf{Example 2:}
When the linearized equation of motion for a perturbation field admits an analytic solution, one can find the spectrum of the normal modes by imposing that the near-boundary coefficient corresponding to the source vanishes. The combination $Z_\eta$ in Example 2 is amenable to such an analytical approach upon a suitable change of radial coordinate. More specifically, in the radial coordinate $r\in\left(-\infty,+\infty\right)$, we find the equation 
\begin{align}\label{eq:Zeta2}
    Z_\eta''
    +
    \frac{3a'}{2a} Z_\eta'
    +
    \left[
    \frac{\omega^2-k^2}{a}
    -\frac{2a\eta_0' V'(\eta_0)}{a'}
    -\frac{a^2\eta_0'^4}{2 a'^2}
    +V''(\eta_0)
    -\frac{3}{2} \eta_0'^2\right]
    Z_\eta
    =     
    0,
\end{align}
where the background solution $\eta_0$ for the field $\eta$ is given in \eqref{eq:sol2} (there, we dropped the subindex 0). For convenience, we have also indicated the warp factor as $a\equiv e^{2A}$. The potential $V$ is derived from the superpotential $W$ given in \eqref{eq:Wex2} according to the general formula \eqref{eq:VfromW},
\begin{eqnarray}
    V(\eta_0)
    =
    -\frac{2\left[2+\cos(\eta_0)\right]}{L^2}\ .
\end{eqnarray}
The equation in the $r$ coordinate is
\begin{equation}
    Z_{\eta }''+\frac{3}{L}
   \tanh \left(\frac{r-r_0}{L}\right)Z_\eta'+\frac{\left(L^2 (\omega^2-k^2)^2
   \operatorname{sech}^2\left(\frac{r-r_0}{L}\right)-8
   \operatorname{csch}^2\left(\frac{r-r_0}{L}\right)+8\right)}{4 L^2} Z_\eta=0\ .
\end{equation} 
In the massless case $\omega^2-k^2=0$, the equations has an analytic solution corresponding to a fluctuation with zero source at both asymptotic boundaries:
\begin{equation}
    Z_\eta(r)=\frac{\alpha_0}{\sinh\left(\frac{r-r_0}{L}\right)}\ ,
\end{equation}
with $\alpha_0$ a constant. It corresponds to a shift of the expectation values of the dual operators in both copies of the field theory, preserving $PT$ symmetry.

Upon compactifying the radial coordinate according to
\begin{eqnarray}
    \label{eq:compa}
    r(v) = r_0 + L \log \left(\frac{1+v}{1-v}\right)
    \ , \qquad 
    z_\eta[v(r)] \equiv Z_\eta\left(r\right)\ ,
\end{eqnarray}
with $v\in\left[-1,1\right]$ and $r_0$, $L>0$ real constants, we obtain the equation 
\begin{eqnarray}
    \label{eq:zeta}
    z_\eta''
    +
    \frac{2 v \left(v^2-5\right) }{v^4-1}
    z_\eta'
    +
    \left[\frac{M^2}{\left(v^2+1\right)^2}-\frac{2 \left(v^4-6 v^2+1\right)}{v^2 \left(v^2-1\right)^2}\right] 
    z_\eta
    =
    0\ ,
\end{eqnarray}
where $M^2/L^2 = \omega^2-k^2$ and the prime denotes the derivative with respect to $v$. This equation admits an analytical solution; we avoid writing it here because it is quite involved%
\footnote{The full solution can be found in \cite{code}.}. It is important to note that the equation \eqref{eq:zeta} is singular at $v=0$, within the domain $v\in\left[-1,1\right]$. Note that the definition of $Z_\eta$ in \eqref{eq:zetaeta} has a $1/A'$ factor that becomes singular at the throat; this corresponds to the $v=0$ singular point. However, although one of the solutions for $Z_\eta$ is singular at $v=0$, both solutions give regular metric components and scalar fluctuations. Thus, one should just require that the coefficients of the Frobenius expansions match when going from $v<0$ to $v>0$. Alternatively, one can use the gauge-invariant field $Y_\eta\equiv A' Z_\eta$.
This happens only for the fluctuation of the ghost scalar; for the fluctuations involving only the metric, the only singular points in the equations are at $v=\pm 1$. In both cases the spectrum splits according to parity under $v\to -v$. For the $Z_\eta$ fluctuation, even modes correspond to the Frobenius solution that vanishes at $v=0$ ($Z_\eta \sim v^2$), while odd modes correspond to the singular solution ($Z_\eta\sim 1/v$). \\

Consider the problem on the negative domain $v<0$. In addition to the boundary condition at $v=0$, we impose that the fall-off associated with the source vanishes at $v=-1$. Since we need to consider alternative quantization, we actually enforce that the ratio of the subleading coefficient over the leading one vanishes in the $r$ coordinate. This allows us to read the values of $M^2$ for which this occurs, namely the squared masses of the normal modes. The results concerning the three lowest-lying modes are reported in Table \ref{tab:spectra}. Given the parity symmetry of the equation \eqref{eq:zeta}, the subproblem concerning the positive $v$ domain is identical to the problem on the negative $v$ domain.
\begin{table}[]
    \centering
    \begin{tabular}{c|c|llll|c}
      & parity & \multicolumn{4}{c|}{$M^2$} &
      \\ \hline
      Example 1 & even & 13.397, & 53.933, &  117.15, &  \dots & \text{(spectral)}
      \\
       & odd &   0, &   30.893, & 82.677, & \dots & \text{(spectral)}
      \\ \hline
      Example 2 & even &  12  ,  & 40, & 160, & \dots & \text{(analytical)}
      \\
       & odd &  0, & 32, & 96, &  \dots & \text{(analytical)}
      \\ \hline
      Example 5 & even & 4.8205, & 36.051, & 40.097, &  \dots & \text{(Wronskian)}
      \\
       & odd & -3.402, & 15.619, & 19.438, & \dots & \text{(Wronskian)}
    \end{tabular}
    \caption{Lowest-lying $M^2$ eigenvalues in the spectrum of normal modes of the ghost scalar fluctuations, for modes even and odd under reflection about the throat of the wormhole. No other eigenvalues were found in the range $-50\leq M^2\leq 130$ (Example 1), $150$ (Example 2) and $140$ (Example 5).}
    \label{tab:spectra}
\end{table}

\noindent 
\textbf{Example 1:}
The equation for $Z_\eta$ in Example 1 is
\begin{align}
    Z_\eta''
    +
    \frac{2a'}{a} Z_\eta'
    +
    \left[
    \frac{\omega^2-k^2}{a}
    -\frac{4 a \eta_0' V'(\eta_0)}{3a'}
    -\frac{2a^2\eta_0'^4}{9a'^2}
    +V''(\eta_0)
    -\frac{4}{3} \eta_0'^2\right]
    Z_\eta
    =     
    0\ ,
\end{align}
where the background profile of the scalar $\eta$ is given in \eqref{eq:sol1} and $a= e^{2A}$. The potential is derived from the superpotential \eqref{eq:Wex1},
\begin{eqnarray}
    V(\eta_0)
    =
    -\frac{3\left[3+\cos\left(\frac{2\eta_0}{\sqrt{3}}\right)\right]}{L^2}\ .
\end{eqnarray}
One can again perform the same change of radial coordinate as in \eqref{eq:compa}, leading to
\begin{align}
    \label{eq:zeta1}
    z_\eta''
    &+
    \frac{2 v \left(v^4-10v^2-15\right) }{v^6+5v^4-5v^2-1}
    z_\eta'
    \nonumber \\ &\qquad +
    2\left[\frac{M^2}{1+6v^2+v^4}
    -
    \frac{v^8-12v^6-10v^4-12v^2+1}{v^2 \left(v^4-1\right)^2}\right] 
    z_\eta
    =
    0,
\end{align}
with $M^2/L^2 = \omega^2 - k^2$.
The properties of the equation are similar to Example 2, with a singular point at $v=0$ and modes split in even and odd under $v\to -v$. As in the previous example, this only happens for the scalar fluctuation.

Apparently, it is not easy to handle equation \eqref{eq:zeta1} analytically. We resort to a spectral method discretized on a Gauss--Lobatto collocation grid and adopt Chebyshev polynomials as the basis functions. Let us focus on the subproblem in the negative $v\in\left[-1,0\right]$ domain. We redefine the fluctuating field $z_\eta$ in such a way that the boundary conditions reduce to the simple requirement of finiteness at the two extrema of the domain. More specifically, the leading and subleading behaviors of the field $z_\eta$ near the singular points $v=-1$ and $v=0$ are, respectively,
\begin{align}
    z_\eta &=
    2(1+v)^2 \log(1+v)\, a_1
    +
    (1+v)^2\, a_2 
    + ...
    \ , \\
    z_\eta &=
    \frac{b_1}{v}
    + v^2\, b_2 + ...\ .
\end{align}
We then define, for the modes,
\begin{equation}
\text{even:}\    z_\eta(v) 
    \equiv 
    v^2 (1-v^2)^2 \psi(v)\;;\quad \text{odd:}\    z_\eta(v) 
    \equiv 
    \frac{1}{v}(1-v^2)^2 \psi(v)  ,
\end{equation}
where we should fix $\psi(v)$ to be even functions. This fixes the parity of the solution and we also factor out explicitly the subleading behaviors at the two extrema of the integration domain. In this way, a finite value for $\psi$ at $v=-1$ and $v=1$ corresponds to a normalizable even or odd mode. 

There is still a technical but important step before actually solving by means of the Chebyshev spectral method. In order to adapt the interval $v\in\left[-1,0\right]$ to the domain $s\in\left[-1,1\right]$ on which the Chebyshev polynomials are defined, we perform a further change of variable $v=-\sqrt{\frac{1-s}{2}}$. After multiplying by the singular factor $(1-s^2)$, the final equation is given by
\begin{eqnarray}
    (1-s^2)\, \psi''(s)
    + 
    \left[A_0(s) + A_1(s) M^2\right]\psi'(s)
    +
    \left[B_0(s) + B_1(s) M^2\right] \psi(s)
    =
    0
\end{eqnarray}
where the prime denotes a derivative with respect to $s$. We have split the coefficient functions $A$ and $B$ according to the order in $M^2$. 

Their explicit expressions, for the even modes, are
\begin{align}
    A_0(s) &\equiv 
    \frac{s ((151-15 s) s-69)-107}{2 (s-14) s+34}
    \ , \\
    A_1(s) &\equiv 0
    \ , \\ 
    B_0(s) &\equiv 
    \frac{2 (s (s ((62-5 s) s-248)+258)+61)}{(s-3)^2 ((s-14) s+17)}\ , \\
    B_1(s) &\equiv 
   \frac{s+1}{(s-14) s+17}\ .
\end{align}

The coefficients for the odds more are
\begin{align}
    A_0(s) &\equiv 
    \frac{s ((73-9 s) s-51)-5}{2 (s-14) s+34}
    \ , \\
    A_1(s) &\equiv 0
    \ , \\ 
    B_0(s) &\equiv -\frac{32 (s-1)}{(s-14)
   s+17}+\frac{4}{s-3}+\frac{16}{(s-3)^2}-\frac{5}{2}\ , \\
    B_1(s) &\equiv 
    \frac{s+1}{(s-14) s+17}\ .
\end{align}

The results are presented in Table \ref{tab:spectra}. The values shown in the table for the lowest-lying eigenvalues $M^2$ are stable upon doubling the number of collocation points from 100 to 200. \\
The $M^2=0$ solution can be found analytically solving \eqref{eq:zeta1}
\begin{equation}
    z_\eta(v)=\frac{\alpha_0  \left(1-v^2\right)^2}{2 v\left(1+v^2\right)}\ ,
\end{equation}
with $\alpha_0$ a constant. It corresponds to a shift of the expectation values of the dual operators in both copies of the field theory, preserving $PT$ symmetry.\\

\noindent 
\textbf{Example 5:} Here, we have two ghost fields, $\eta_{1,2}$, and we can construct the associated gauge-invariant combinations according to \eqref{eq:zetaeta}; let us denote the latter by $Z_{1,2}$. Although $Z_{1,2}$ do not source the gauge-invariant modes $H$ of the metric, still $Z_1$ and $Z_2$ are coupled to each other. 

Since this case presents cumbersome expressions, we leave the explicit details about the equations and their handling to the notebook shared at \cite{code}.  We tackled the study of the spectrum with a numerical shooting method. The shooting method entails finding an asymptotic series expansion at each of the two extrema of the integration domain. Again, as in Examples 1 and 2,  the equations have a singular point at the throat of the wormhole (the center of the bulk), but we can split the spectrum into even and odd modes corresponding to a specific choice of the solution at the singular point. The fluctuation problem can then be solved in the region extending from the throat to the corresponding boundary. Other details are different, since we are dealing with a system of two coupled fields, rather than a single equation. One needs to enforce the desired boundary conditions on these expansions and then feed them as initial conditions to a numerical solver. One thus finds a numerical solution propagating from the boundary towards the bulk interior and another numerical solution propagating from the throat towards the boundary.

The two numerical solutions are then matched at an intermediate point. Specifically, one asks that the Wronskian built from the numerical solutions and their derivatives evaluated at the matching point vanishes. This means that the two solutions are linearly dependent (since the problem is linear, they are actually the same solution). The values of $M^2$ at which the Wronskian vanishes are the sought-for eigenvalues. The results for the lowest modes are also listed in Table \ref{tab:spectra}.\\

\bibliographystyle{JHEP}
\bibliography{Refs}

\providecommand{\href}[2]{#2}\begingroup\raggedright\begin{thebibliography}{100}

\bibitem{Witten:2010cx}
E.~Witten, \emph{{Analytic Continuation Of Chern-Simons Theory}}, {\emph{AMS/IP
  Stud. Adv. Math.} {\bfseries 50} (2011) 347}
  [\href{https://arxiv.org/abs/1001.2933}{{\ttfamily 1001.2933}}].

\bibitem{Basar:2013eka}
G.~Basar, G.~V. Dunne and M.~Unsal, \emph{{Resurgence theory, ghost-instantons,
  and analytic continuation of path integrals}},
  \href{https://doi.org/10.1007/JHEP10(2013)041}{\emph{JHEP} {\bfseries 10}
  (2013) 041} [\href{https://arxiv.org/abs/1308.1108}{{\ttfamily 1308.1108}}].

\bibitem{Gorbenko:2018dtm}
V.~Gorbenko, S.~Rychkov and B.~Zan, \emph{{Walking, Weak first-order
  transitions, and Complex CFTs II. Two-dimensional Potts model at $Q>4$}},
  \href{https://doi.org/10.21468/SciPostPhys.5.5.050}{\emph{SciPost Phys.}
  {\bfseries 5} (2018) 050} [\href{https://arxiv.org/abs/1808.04380}{{\ttfamily
  1808.04380}}].

\bibitem{Gorbenko:2018ncu}
V.~Gorbenko, S.~Rychkov and B.~Zan, \emph{{Walking, Weak first-order
  transitions, and Complex CFTs}},
  \href{https://doi.org/10.1007/JHEP10(2018)108}{\emph{JHEP} {\bfseries 10}
  (2018) 108} [\href{https://arxiv.org/abs/1807.11512}{{\ttfamily
  1807.11512}}].

\bibitem{Mostafazadeh:2008pw}
A.~Mostafazadeh, \emph{{Pseudo-Hermitian Representation of Quantum Mechanics}},
  \href{https://doi.org/10.1142/S0219887810004816}{\emph{Int. J. Geom. Meth.
  Mod. Phys.} {\bfseries 7} (2010) 1191}
  [\href{https://arxiv.org/abs/0810.5643}{{\ttfamily 0810.5643}}].

\bibitem{Ashida:2020dkc}
Y.~Ashida, Z.~Gong and M.~Ueda, \emph{{Non-Hermitian physics}},
  \href{https://doi.org/10.1080/00018732.2021.1876991}{\emph{Adv. Phys.}
  {\bfseries 69} (2021) 249}
  [\href{https://arxiv.org/abs/2006.01837}{{\ttfamily 2006.01837}}].

\bibitem{Bender:2023cem}
C.~M. Bender and D.~W. Hook, \emph{{PT-symmetric quantum mechanics}},
  \href{https://arxiv.org/abs/2312.17386}{{\ttfamily 2312.17386}}.

\bibitem{Faedo:2019nxw}
A.~F. Faedo, C.~Hoyos, D.~Mateos and J.~G. Subils, \emph{{Holographic Complex
  Conformal Field Theories}},
  \href{https://doi.org/10.1103/PhysRevLett.124.161601}{\emph{Phys. Rev. Lett.}
  {\bfseries 124} (2020) 161601}
  [\href{https://arxiv.org/abs/1909.04008}{{\ttfamily 1909.04008}}].

\bibitem{Faedo:2021ksi}
A.~F. Faedo, C.~Hoyos, D.~Mateos and J.~G. Subils, \emph{{Multiple mass
  hierarchies from complex fixed point collisions}},
  \href{https://doi.org/10.1007/JHEP10(2021)246}{\emph{JHEP} {\bfseries 10}
  (2021) 246} [\href{https://arxiv.org/abs/2106.01802}{{\ttfamily
  2106.01802}}].

\bibitem{Arean:2019pom}
D.~Are{\'a}n, K.~Landsteiner and I.~Salazar~Landea, \emph{{Non-hermitian
  holography}},
  \href{https://doi.org/10.21468/SciPostPhys.9.3.032}{\emph{SciPost Phys.}
  {\bfseries 9} (2020) 032} [\href{https://arxiv.org/abs/1912.06647}{{\ttfamily
  1912.06647}}].

\bibitem{Morales-Tejera:2022hyq}
S.~Morales-Tejera and K.~Landsteiner, \emph{{Non-Hermitian quantum quenches in
  holography}},
  \href{https://doi.org/10.21468/SciPostPhys.14.3.030}{\emph{SciPost Phys.}
  {\bfseries 14} (2023) 030}
  [\href{https://arxiv.org/abs/2203.02524}{{\ttfamily 2203.02524}}].

\bibitem{Xian:2023zgu}
Z.-Y. Xian, D.~Rodr{\'\i}guez~Fern{\'a}ndez, Z.~Chen, Y.~Liu and R.~Meyer,
  \emph{{Electric conductivity in non-Hermitian holography}},
  \href{https://doi.org/10.21468/SciPostPhys.16.1.004}{\emph{SciPost Phys.}
  {\bfseries 16} (2024) 004}
  [\href{https://arxiv.org/abs/2304.11183}{{\ttfamily 2304.11183}}].

\bibitem{Arean:2024gks}
D.~Arean, D.~Garcia-Fari{\~n}a and K.~Landsteiner, \emph{{Strongly Coupled
  PT-Symmetric Models in Holography}},
  \href{https://doi.org/10.3390/e27010013}{\emph{Entropy} {\bfseries 27} (2025)
  13} [\href{https://arxiv.org/abs/2411.18471}{{\ttfamily 2411.18471}}].

\bibitem{Arean:2024lzz}
D.~Arean and D.~Garcia-Fari{\~n}a, \emph{{Holographic non-Hermitian lattices
  and junctions and their RG flows}},
  \href{https://doi.org/10.1007/JHEP07(2025)276}{\emph{JHEP} {\bfseries 07}
  (2025) 276} [\href{https://arxiv.org/abs/2410.13584}{{\ttfamily
  2410.13584}}].

\bibitem{Maldacena:2018lmt}
J.~Maldacena and X.-L. Qi, \emph{{Eternal traversable wormhole}},
  \href{https://arxiv.org/abs/1804.00491}{{\ttfamily 1804.00491}}.

\bibitem{Loges:2022nuw}
G.~J. Loges, G.~Shiu and N.~Sudhir, \emph{{Complex saddles and Euclidean
  wormholes in the Lorentzian path integral}},
  \href{https://doi.org/10.1007/JHEP08(2022)064}{\emph{JHEP} {\bfseries 08}
  (2022) 064} [\href{https://arxiv.org/abs/2203.01956}{{\ttfamily
  2203.01956}}].

\bibitem{Garcia-Garcia:2020ttf}
A.~M. Garc{\'\i}a-Garc{\'\i}a and V.~Godet, \emph{{Euclidean wormhole in the
  Sachdev-Ye-Kitaev model}},
  \href{https://doi.org/10.1103/PhysRevD.103.046014}{\emph{Phys. Rev. D}
  {\bfseries 103} (2021) 046014}
  [\href{https://arxiv.org/abs/2010.11633}{{\ttfamily 2010.11633}}].

\bibitem{Garcia-Garcia:2022zmo}
A.~M. Garc{\'\i}a-Garc{\'\i}a, V.~Godet, C.~Yin and J.~P. Zheng,
  \emph{{Euclidean-to-Lorentzian wormhole transition and gravitational symmetry
  breaking in the Sachdev-Ye-Kitaev model}},
  \href{https://doi.org/10.1103/PhysRevD.106.046008}{\emph{Phys. Rev. D}
  {\bfseries 106} (2022) 046008}
  [\href{https://arxiv.org/abs/2204.08558}{{\ttfamily 2204.08558}}].

\bibitem{Harper:2025lav}
J.~Harper, T.~Kawamoto, R.~Maeda, N.~Nakamura and T.~Takayanagi,
  \emph{{Non-hermitian Density Matrices from Time-like Entanglement and
  Wormholes}},  \href{https://arxiv.org/abs/2512.13800}{{\ttfamily
  2512.13800}}.

\bibitem{Kawamoto:2025oko}
T.~Kawamoto, R.~Maeda, N.~Nakamura and T.~Takayanagi, \emph{{Traversable AdS
  wormhole via non-local double trace or Janus deformation}},
  \href{https://doi.org/10.1007/JHEP04(2025)086}{\emph{JHEP} {\bfseries 04}
  (2025) 086} [\href{https://arxiv.org/abs/2502.03531}{{\ttfamily
  2502.03531}}].

\bibitem{Held:2026bbo}
J.~Held, M.~Kaplan, D.~Marolf and Z.~Wang, \emph{{Lorentzian Path Integrals and
  Jackiw-Teitelboim wormholes with imaginary scalars}},
  \href{https://arxiv.org/abs/2601.09932}{{\ttfamily 2601.09932}}.

\bibitem{Gibbons:1978ac}
G.~W. Gibbons, S.~W. Hawking and M.~J. Perry, \emph{{Path Integrals and the
  Indefiniteness of the Gravitational Action}},
  \href{https://doi.org/10.1016/0550-3213(78)90161-X}{\emph{Nucl. Phys. B}
  {\bfseries 138} (1978) 141}.

\bibitem{Halliwell:1989dy}
J.~J. Halliwell and J.~B. Hartle, \emph{{Integration Contours for the No
  Boundary Wave Function of the Universe}},
  \href{https://doi.org/10.1103/PhysRevD.41.1815}{\emph{Phys. Rev. D}
  {\bfseries 41} (1990) 1815}.

\bibitem{Louko:1995jw}
J.~Louko and R.~D. Sorkin, \emph{{Complex actions in two-dimensional topology
  change}}, \href{https://doi.org/10.1088/0264-9381/14/1/018}{\emph{Class.
  Quant. Grav.} {\bfseries 14} (1997) 179}
  [\href{https://arxiv.org/abs/gr-qc/9511023}{{\ttfamily gr-qc/9511023}}].

\bibitem{Nekrasov:2023xzm}
N.~Nekrasov, \emph{{Analytic continuation and supersymmetry.}},
  \href{https://doi.org/10.1090/pspum/107/01950}{\emph{Proc. Symp. Pure Math.}
  {\bfseries 107} (2024) 167}
  [\href{https://arxiv.org/abs/2310.01654}{{\ttfamily 2310.01654}}].

\bibitem{Zigdon:2026xal}
Y.~Zigdon, \emph{{Bridging Worldsheet CFTs and Wormholes}},
  \href{https://arxiv.org/abs/2603.15739}{{\ttfamily 2603.15739}}.

\bibitem{Kontsevich:2021dmb}
M.~Kontsevich and G.~Segal, \emph{{Wick Rotation and the Positivity of Energy
  in Quantum Field Theory}},
  \href{https://doi.org/10.1093/qmath/haab027}{\emph{Quart. J. Math. Oxford
  Ser.} {\bfseries 72} (2021) 673}
  [\href{https://arxiv.org/abs/2105.10161}{{\ttfamily 2105.10161}}].

\bibitem{Witten:2021nzp}
E.~Witten, \emph{{A Note On Complex Spacetime Metrics}},
  \href{https://arxiv.org/abs/2111.06514}{{\ttfamily 2111.06514}}.

\bibitem{Lehners:2021mah}
J.-L. Lehners, \emph{{Allowable complex metrics in minisuperspace quantum
  cosmology}}, \href{https://doi.org/10.1103/PhysRevD.105.026022}{\emph{Phys.
  Rev. D} {\bfseries 105} (2022) 026022}
  [\href{https://arxiv.org/abs/2111.07816}{{\ttfamily 2111.07816}}].

\bibitem{Krishna:2026rma}
V.~Krishna and F.~Larsen, \emph{{Allowable Complex Black Holes in the Euclidean
  Gravitational Path Integral}},
  \href{https://arxiv.org/abs/2602.05979}{{\ttfamily 2602.05979}}.

\bibitem{Bergshoeff:2007cg}
E.~A. Bergshoeff, J.~Hartong, A.~Ploegh, J.~Rosseel and D.~Van~den Bleeken,
  \emph{{Pseudo-supersymmetry and a tale of alternate realities}},
  \href{https://doi.org/10.1088/1126-6708/2007/07/067}{\emph{JHEP} {\bfseries
  07} (2007) 067} [\href{https://arxiv.org/abs/0704.3559}{{\ttfamily
  0704.3559}}].

\bibitem{Skenderis:2007sm}
K.~Skenderis, P.~K. Townsend and A.~Van~Proeyen, \emph{{Domain-wall/cosmology
  correspondence in adS/dS supergravity}},
  \href{https://doi.org/10.1088/1126-6708/2007/08/036}{\emph{JHEP} {\bfseries
  08} (2007) 036} [\href{https://arxiv.org/abs/0704.3918}{{\ttfamily
  0704.3918}}].

\bibitem{Skenderis:2006jq}
K.~Skenderis and P.~K. Townsend, \emph{{Hidden supersymmetry of domain walls
  and cosmologies}},
  \href{https://doi.org/10.1103/PhysRevLett.96.191301}{\emph{Phys. Rev. Lett.}
  {\bfseries 96} (2006) 191301}
  [\href{https://arxiv.org/abs/hep-th/0602260}{{\ttfamily hep-th/0602260}}].

\bibitem{Giddings:1987cg}
S.~B. Giddings and A.~Strominger, \emph{{Axion Induced Topology Change in
  Quantum Gravity and String Theory}},
  \href{https://doi.org/10.1016/0550-3213(88)90446-4}{\emph{Nucl. Phys. B}
  {\bfseries 306} (1988) 890}.

\bibitem{Gutperle:2002km}
M.~Gutperle and W.~Sabra, \emph{{Instantons and wormholes in Minkowski and
  (A)dS spaces}},
  \href{https://doi.org/10.1016/S0550-3213(02)00942-2}{\emph{Nucl. Phys. B}
  {\bfseries 647} (2002) 344}
  [\href{https://arxiv.org/abs/hep-th/0206153}{{\ttfamily hep-th/0206153}}].

\bibitem{Bergshoeff:2004pg}
E.~Bergshoeff, A.~Collinucci, U.~Gran, D.~Roest and S.~Vandoren,
  \emph{{Non-extremal instantons and wormholes in string theory}},
  \href{https://doi.org/10.1002/prop.200410227}{\emph{Fortsch. Phys.}
  {\bfseries 53} (2005) 990}
  [\href{https://arxiv.org/abs/hep-th/0412183}{{\ttfamily hep-th/0412183}}].

\bibitem{Bergshoeff:2005zf}
E.~Bergshoeff, A.~Collinucci, A.~Ploegh, S.~Vandoren and T.~Van~Riet,
  \emph{{Non-extremal D-instantons and the AdS/CFT correspondence}},
  \href{https://doi.org/10.1088/1126-6708/2006/01/061}{\emph{JHEP} {\bfseries
  01} (2006) 061} [\href{https://arxiv.org/abs/hep-th/0510048}{{\ttfamily
  hep-th/0510048}}].

\bibitem{Arkani-Hamed:2007cpn}
N.~Arkani-Hamed, J.~Orgera and J.~Polchinski, \emph{{Euclidean wormholes in
  string theory}},
  \href{https://doi.org/10.1088/1126-6708/2007/12/018}{\emph{JHEP} {\bfseries
  12} (2007) 018} [\href{https://arxiv.org/abs/0705.2768}{{\ttfamily
  0705.2768}}].

\bibitem{Bergshoeff:2008be}
E.~Bergshoeff, W.~Chemissany, A.~Ploegh, M.~Trigiante and T.~Van~Riet,
  \emph{{Generating Geodesic Flows and Supergravity Solutions}},
  \href{https://doi.org/10.1016/j.nuclphysb.2008.10.023}{\emph{Nucl. Phys. B}
  {\bfseries 812} (2009) 343}
  [\href{https://arxiv.org/abs/0806.2310}{{\ttfamily 0806.2310}}].

\bibitem{Hertog:2017owm}
T.~Hertog, M.~Trigiante and T.~Van~Riet, \emph{{Axion Wormholes in AdS
  Compactifications}},
  \href{https://doi.org/10.1007/JHEP06(2017)067}{\emph{JHEP} {\bfseries 06}
  (2017) 067} [\href{https://arxiv.org/abs/1702.04622}{{\ttfamily
  1702.04622}}].

\bibitem{Ruggeri:2017grz}
D.~Ruggeri, M.~Trigiante and T.~Van~Riet, \emph{{Instantons from geodesics in
  AdS moduli spaces}},
  \href{https://doi.org/10.1007/JHEP03(2018)091}{\emph{JHEP} {\bfseries 03}
  (2018) 091} [\href{https://arxiv.org/abs/1712.06081}{{\ttfamily
  1712.06081}}].

\bibitem{Marolf:2021kjc}
D.~Marolf and J.~E. Santos, \emph{{AdS Euclidean wormholes}},
  \href{https://doi.org/10.1088/1361-6382/ac2cb7}{\emph{Class. Quant. Grav.}
  {\bfseries 38} (2021) 224002}
  [\href{https://arxiv.org/abs/2101.08875}{{\ttfamily 2101.08875}}].

\bibitem{Astesiano:2022qba}
D.~Astesiano, D.~Ruggeri, M.~Trigiante and T.~Van~Riet, \emph{{Instantons and
  no wormholes in $AdS_3\times S^3 \times CY_2$}},
  \href{https://doi.org/10.1103/PhysRevD.105.086022}{\emph{Phys. Rev. D}
  {\bfseries 105} (2022) 086022}
  [\href{https://arxiv.org/abs/2201.11694}{{\ttfamily 2201.11694}}].

\bibitem{Loges:2023ypl}
G.~J. Loges, G.~Shiu and T.~Van~Riet, \emph{{A 10d construction of Euclidean
  axion wormholes in flat and AdS space}},
  \href{https://doi.org/10.1007/JHEP06(2023)079}{\emph{JHEP} {\bfseries 06}
  (2023) 079} [\href{https://arxiv.org/abs/2302.03688}{{\ttfamily
  2302.03688}}].

\bibitem{Astesiano:2023iql}
D.~Astesiano and F.~F. Gautason, \emph{{Supersymmetric Wormholes in String
  Theory}}, \href{https://doi.org/10.1103/PhysRevLett.132.161601}{\emph{Phys.
  Rev. Lett.} {\bfseries 132} (2024) 161601}
  [\href{https://arxiv.org/abs/2309.02481}{{\ttfamily 2309.02481}}].

\bibitem{Anabalon:2023kcp}
A.~Anabal{\'o}n, {\'A}.~Arboleya and A.~Guarino, \emph{{Euclidean flows,
  solitons, and wormholes in AdS space from M-theory}},
  \href{https://doi.org/10.1103/PhysRevD.109.106007}{\emph{Phys. Rev. D}
  {\bfseries 109} (2024) 106007}
  [\href{https://arxiv.org/abs/2312.13955}{{\ttfamily 2312.13955}}].

\bibitem{DiUbaldo:2026rly}
G.~Di~Ubaldo, L.~V. Iliesiu, H.~W. Lin and C.~Yan, \emph{{Positivity of the
  gravitational path integral implies the axionic weak gravity conjecture}},
  \href{https://arxiv.org/abs/2605.05305}{{\ttfamily 2605.05305}}.

\bibitem{Hebecker:2018ofv}
A.~Hebecker, T.~Mikhail and P.~Soler, \emph{{Euclidean wormholes, baby
  universes, and their impact on particle physics and cosmology}},
  \href{https://doi.org/10.3389/fspas.2018.00035}{\emph{Front. Astron. Space
  Sci.} {\bfseries 5} (2018) 35}
  [\href{https://arxiv.org/abs/1807.00824}{{\ttfamily 1807.00824}}].

\bibitem{Maldacena:2026jqd}
J.~Maldacena, A.~Maloney and B.~McPeak, \emph{{Wormholes and the imaginary
  distance bound}},  \href{https://arxiv.org/abs/2605.05336}{{\ttfamily
  2605.05336}}.

\bibitem{Freedman:1999gk}
D.~Z. Freedman, S.~S. Gubser, K.~Pilch and N.~P. Warner, \emph{{Continuous
  distributions of D3-branes and gauged supergravity}},
  \href{https://doi.org/10.1088/1126-6708/2000/07/038}{\emph{JHEP} {\bfseries
  07} (2000) 038} [\href{https://arxiv.org/abs/hep-th/9906194}{{\ttfamily
  hep-th/9906194}}].

\bibitem{Cvetic:1999xx}
M.~Cvetic, S.~S. Gubser, H.~Lu and C.~N. Pope, \emph{{Symmetric potentials of
  gauged supergravities in diverse dimensions and Coulomb branch of gauge
  theories}}, \href{https://doi.org/10.1103/PhysRevD.62.086003}{\emph{Phys.
  Rev. D} {\bfseries 62} (2000) 086003}
  [\href{https://arxiv.org/abs/hep-th/9909121}{{\ttfamily hep-th/9909121}}].

\bibitem{Tarrio:2013qga}
J.~Tarr{\'\i}o and O.~Varela, \emph{{Electric/magnetic duality and RG flows in
  AdS$_4$/CFT$_3$}}, \href{https://doi.org/10.1007/JHEP01(2014)071}{\emph{JHEP}
  {\bfseries 01} (2014) 071} [\href{https://arxiv.org/abs/1311.2933}{{\ttfamily
  1311.2933}}].

\bibitem{Pilch:2000ue}
K.~Pilch and N.~P. Warner, \emph{{N=2 supersymmetric RG flows and the IIB
  dilaton}}, \href{https://doi.org/10.1016/S0550-3213(00)00656-8}{\emph{Nucl.
  Phys. B} {\bfseries 594} (2001) 209}
  [\href{https://arxiv.org/abs/hep-th/0004063}{{\ttfamily hep-th/0004063}}].

\bibitem{Bobev:2010de}
N.~Bobev, A.~Kundu, K.~Pilch and N.~P. Warner, \emph{{Supersymmetric Charged
  Clouds in $AdS_{5}$}},
  \href{https://doi.org/10.1007/JHEP03(2011)070}{\emph{JHEP} {\bfseries 03}
  (2011) 070} [\href{https://arxiv.org/abs/1005.3552}{{\ttfamily 1005.3552}}].

\bibitem{Freedman:1999gp}
D.~Z. Freedman, S.~S. Gubser, K.~Pilch and N.~P. Warner, \emph{{Renormalization
  group flows from holography supersymmetry and a c theorem}},
  \href{https://doi.org/10.4310/ATMP.1999.v3.n2.a7}{\emph{Adv. Theor. Math.
  Phys.} {\bfseries 3} (1999) 363}
  [\href{https://arxiv.org/abs/hep-th/9904017}{{\ttfamily hep-th/9904017}}].

\bibitem{Pilch:2000fu}
K.~Pilch and N.~P. Warner, \emph{{N=1 supersymmetric renormalization group
  flows from IIB supergravity}},
  \href{https://doi.org/10.4310/ATMP.2000.v4.n3.a5}{\emph{Adv. Theor. Math.
  Phys.} {\bfseries 4} (2002) 627}
  [\href{https://arxiv.org/abs/hep-th/0006066}{{\ttfamily hep-th/0006066}}].

\bibitem{Girardello:1999bd}
L.~Girardello, M.~Petrini, M.~Porrati and A.~Zaffaroni, \emph{{The Supergravity
  dual of N=1 superYang-Mills theory}},
  \href{https://doi.org/10.1016/S0550-3213(99)00764-6}{\emph{Nucl. Phys. B}
  {\bfseries 569} (2000) 451}
  [\href{https://arxiv.org/abs/hep-th/9909047}{{\ttfamily hep-th/9909047}}].

\bibitem{Bobev:2018eer}
N.~Bobev, F.~F. Gautason, B.~E. Niehoff and J.~van Muiden, \emph{{Uplifting
  GPPZ: a ten-dimensional dual of $ \mathcal{N}={1}^{\ast } $}},
  \href{https://doi.org/10.1007/JHEP10(2018)058}{\emph{JHEP} {\bfseries 10}
  (2018) 058} [\href{https://arxiv.org/abs/1805.03623}{{\ttfamily
  1805.03623}}].

\bibitem{Petrini:2018pjk}
M.~Petrini, H.~Samtleben, S.~Schmidt and K.~Skenderis, \emph{{The 10d Uplift of
  the GPPZ Solution}},
  \href{https://doi.org/10.1007/JHEP07(2018)026}{\emph{JHEP} {\bfseries 07}
  (2018) 026} [\href{https://arxiv.org/abs/1805.01919}{{\ttfamily
  1805.01919}}].

\bibitem{Gubser:2009qm}
S.~S. Gubser, C.~P. Herzog, S.~S. Pufu and T.~Tesileanu, \emph{{Superconductors
  from Superstrings}},
  \href{https://doi.org/10.1103/PhysRevLett.103.141601}{\emph{Phys. Rev. Lett.}
  {\bfseries 103} (2009) 141601}
  [\href{https://arxiv.org/abs/0907.3510}{{\ttfamily 0907.3510}}].

\bibitem{Cassani:2010uw}
D.~Cassani, G.~Dall'Agata and A.~F. Faedo, \emph{{Type IIB supergravity on
  squashed Sasaki-Einstein manifolds}},
  \href{https://doi.org/10.1007/JHEP05(2010)094}{\emph{JHEP} {\bfseries 05}
  (2010) 094} [\href{https://arxiv.org/abs/1003.4283}{{\ttfamily 1003.4283}}].

\bibitem{Gauntlett:2010vu}
J.~P. Gauntlett and O.~Varela, \emph{{Universal Kaluza-Klein reductions of type
  IIB to N=4 supergravity in five dimensions}},
  \href{https://doi.org/10.1007/JHEP06(2010)081}{\emph{JHEP} {\bfseries 06}
  (2010) 081} [\href{https://arxiv.org/abs/1003.5642}{{\ttfamily 1003.5642}}].

\bibitem{Gauntlett:2009dn}
J.~P. Gauntlett, J.~Sonner and T.~Wiseman, \emph{{Holographic superconductivity
  in M-Theory}},
  \href{https://doi.org/10.1103/PhysRevLett.103.151601}{\emph{Phys. Rev. Lett.}
  {\bfseries 103} (2009) 151601}
  [\href{https://arxiv.org/abs/0907.3796}{{\ttfamily 0907.3796}}].

\bibitem{Huang:2020qmn}
H.~Huang, H.~L{\"u} and J.~Yang, \emph{{Bronnikov-like wormholes in
  Einstein-scalar gravity}},
  \href{https://doi.org/10.1088/1361-6382/ac8266}{\emph{Class. Quant. Grav.}
  {\bfseries 39} (2022) 185009}
  [\href{https://arxiv.org/abs/2010.00197}{{\ttfamily 2010.00197}}].

\bibitem{Nozawa:2020gzz}
M.~Nozawa, \emph{{Static spacetimes haunted by a phantom scalar field III:
  asymptotically (A)dS solutions}},
  \href{https://doi.org/10.1103/PhysRevD.103.024005}{\emph{Phys. Rev. D}
  {\bfseries 103} (2021) 024005}
  [\href{https://arxiv.org/abs/2010.07561}{{\ttfamily 2010.07561}}].

\bibitem{Morris:1988tu}
M.~S. Morris, K.~S. Thorne and U.~Yurtsever, \emph{{Wormholes, Time Machines,
  and the Weak Energy Condition}},
  \href{https://doi.org/10.1103/PhysRevLett.61.1446}{\emph{Phys. Rev. Lett.}
  {\bfseries 61} (1988) 1446}.

\bibitem{Hochberg:1998ii}
D.~Hochberg and M.~Visser, \emph{{The Null energy condition in dynamic
  wormholes}}, \href{https://doi.org/10.1103/PhysRevLett.81.746}{\emph{Phys.
  Rev. Lett.} {\bfseries 81} (1998) 746}
  [\href{https://arxiv.org/abs/gr-qc/9802048}{{\ttfamily gr-qc/9802048}}].

\bibitem{Hoyos:2010at}
C.~Hoyos and P.~Koroteev, \emph{{On the Null Energy Condition and Causality in
  Lifshitz Holography}},
  \href{https://doi.org/10.1103/PhysRevD.82.109905}{\emph{Phys. Rev. D}
  {\bfseries 82} (2010) 084002}
  [\href{https://arxiv.org/abs/1007.1428}{{\ttfamily 1007.1428}}].

\bibitem{Gibbons:1983aq}
G.~W. Gibbons, C.~M. Hull and N.~P. Warner, \emph{{The Stability of Gauged
  Supergravity}},
  \href{https://doi.org/10.1016/0550-3213(83)90480-7}{\emph{Nucl. Phys. B}
  {\bfseries 218} (1983) 173}.

\bibitem{Hull:1983ap}
C.~M. Hull, \emph{{The Positivity of Gravitational Energy and Global
  Supersymmetry}}, \href{https://doi.org/10.1007/BF01216185}{\emph{Commun.
  Math. Phys.} {\bfseries 90} (1983) 545}.

\bibitem{Townsend:1984iu}
P.~K. Townsend, \emph{{Positive Energy and the Scalar Potential in Higher
  Dimensional (Super)gravity Theories}},
  \href{https://doi.org/10.1016/0370-2693(84)91610-1}{\emph{Phys. Lett. B}
  {\bfseries 148} (1984) 55}.

\bibitem{ErrastiDiez:2024hfq}
V.~Errasti~D{\'\i}ez, J.~Gaset~Rif{\`a} and G.~Staudt, \emph{{Foundations of
  Ghost Stability}},
  \href{https://doi.org/10.1002/prop.202400268}{\emph{Fortsch. Phys.}
  {\bfseries 73} (2025) 2400268}
  [\href{https://arxiv.org/abs/2408.16832}{{\ttfamily 2408.16832}}].

\bibitem{Sakai:2004cn}
T.~Sakai and S.~Sugimoto, \emph{{Low energy hadron physics in holographic
  QCD}}, \href{https://doi.org/10.1143/PTP.113.843}{\emph{Prog. Theor. Phys.}
  {\bfseries 113} (2005) 843}
  [\href{https://arxiv.org/abs/hep-th/0412141}{{\ttfamily hep-th/0412141}}].

\bibitem{Casero:2007ae}
R.~Casero, E.~Kiritsis and A.~Paredes, \emph{{Chiral symmetry breaking as open
  string tachyon condensation}},
  \href{https://doi.org/10.1016/j.nuclphysb.2007.07.009}{\emph{Nucl. Phys. B}
  {\bfseries 787} (2007) 98}
  [\href{https://arxiv.org/abs/hep-th/0702155}{{\ttfamily hep-th/0702155}}].

\bibitem{Dhar:2007bz}
A.~Dhar and P.~Nag, \emph{{Sakai-Sugimoto model, Tachyon Condensation and
  Chiral symmetry Breaking}},
  \href{https://doi.org/10.1088/1126-6708/2008/01/055}{\emph{JHEP} {\bfseries
  01} (2008) 055} [\href{https://arxiv.org/abs/0708.3233}{{\ttfamily
  0708.3233}}].

\bibitem{Balasubramanian:2020ffd}
V.~Balasubramanian, M.~Decross and G.~S{\'a}rosi, \emph{{Knitting Wormholes by
  Entanglement in Supergravity}},
  \href{https://doi.org/10.1007/JHEP11(2020)167}{\emph{JHEP} {\bfseries 11}
  (2020) 167} [\href{https://arxiv.org/abs/2009.08980}{{\ttfamily
  2009.08980}}].

\bibitem{Sen:2003tm}
A.~Sen, \emph{{Dirac-Born-Infeld action on the tachyon kink and vortex}},
  \href{https://doi.org/10.1103/PhysRevD.68.066008}{\emph{Phys. Rev. D}
  {\bfseries 68} (2003) 066008}
  [\href{https://arxiv.org/abs/hep-th/0303057}{{\ttfamily hep-th/0303057}}].

\bibitem{Garousi:2004rd}
M.~R. Garousi, \emph{{D-brane anti-D-brane effective action and brane
  interaction in open string channel}},
  \href{https://doi.org/10.1088/1126-6708/2005/01/029}{\emph{JHEP} {\bfseries
  01} (2005) 029} [\href{https://arxiv.org/abs/hep-th/0411222}{{\ttfamily
  hep-th/0411222}}].

\bibitem{Kraus:2000nj}
P.~Kraus and F.~Larsen, \emph{{Boundary string field theory of the D anti-D
  system}}, \href{https://doi.org/10.1103/PhysRevD.63.106004}{\emph{Phys. Rev.
  D} {\bfseries 63} (2001) 106004}
  [\href{https://arxiv.org/abs/hep-th/0012198}{{\ttfamily hep-th/0012198}}].

\bibitem{Takayanagi:2000rz}
T.~Takayanagi, S.~Terashima and T.~Uesugi, \emph{{Brane - anti-brane action
  from boundary string field theory}},
  \href{https://doi.org/10.1088/1126-6708/2001/03/019}{\emph{JHEP} {\bfseries
  03} (2001) 019} [\href{https://arxiv.org/abs/hep-th/0012210}{{\ttfamily
  hep-th/0012210}}].

\bibitem{Jones:2002sia}
N.~T. Jones and S.~H.~H. Tye, \emph{{An Improved brane anti-brane action from
  boundary superstring field theory and multivortex solutions}},
  \href{https://doi.org/10.1088/1126-6708/2003/01/012}{\emph{JHEP} {\bfseries
  01} (2003) 012} [\href{https://arxiv.org/abs/hep-th/0211180}{{\ttfamily
  hep-th/0211180}}].

\bibitem{Jatkar:2000ei}
D.~P. Jatkar, G.~Mandal and S.~R. Wadia, \emph{{Nielsen-Olesen vortices in
  noncommutative Abelian Higgs model}},
  \href{https://doi.org/10.1088/1126-6708/2000/09/018}{\emph{JHEP} {\bfseries
  09} (2000) 018} [\href{https://arxiv.org/abs/hep-th/0007078}{{\ttfamily
  hep-th/0007078}}].

\bibitem{Hikida:2000cp}
Y.~Hikida, M.~Nozaki and T.~Takayanagi, \emph{{Tachyon condensation on fuzzy
  sphere and noncommutative solitons}},
  \href{https://doi.org/10.1016/S0550-3213(00)00693-3}{\emph{Nucl. Phys. B}
  {\bfseries 595} (2001) 319}
  [\href{https://arxiv.org/abs/hep-th/0008023}{{\ttfamily hep-th/0008023}}].

\bibitem{Maldacena:2001kr}
J.~M. Maldacena, \emph{{Eternal black holes in anti-de Sitter}},
  \href{https://doi.org/10.1088/1126-6708/2003/04/021}{\emph{JHEP} {\bfseries
  04} (2003) 021} [\href{https://arxiv.org/abs/hep-th/0106112}{{\ttfamily
  hep-th/0106112}}].

\bibitem{Gao:2016bin}
P.~Gao, D.~L. Jafferis and A.~C. Wall, \emph{{Traversable Wormholes via a
  Double Trace Deformation}},
  \href{https://doi.org/10.1007/JHEP12(2017)151}{\emph{JHEP} {\bfseries 12}
  (2017) 151} [\href{https://arxiv.org/abs/1608.05687}{{\ttfamily
  1608.05687}}].

\bibitem{Witten:2001ua}
E.~Witten, \emph{{Multitrace operators, boundary conditions, and AdS / CFT
  correspondence}},  \href{https://arxiv.org/abs/hep-th/0112258}{{\ttfamily
  hep-th/0112258}}.

\bibitem{Maldacena:2017axo}
J.~Maldacena, D.~Stanford and Z.~Yang, \emph{{Diving into traversable
  wormholes}}, \href{https://doi.org/10.1002/prop.201700034}{\emph{Fortsch.
  Phys.} {\bfseries 65} (2017) 1700034}
  [\href{https://arxiv.org/abs/1704.05333}{{\ttfamily 1704.05333}}].

\bibitem{Caceres:2018ehr}
E.~Caceres, A.~S. Misobuchi and M.-L. Xiao, \emph{{Rotating traversable
  wormholes in AdS}},
  \href{https://doi.org/10.1007/JHEP12(2018)005}{\emph{JHEP} {\bfseries 12}
  (2018) 005} [\href{https://arxiv.org/abs/1807.07239}{{\ttfamily
  1807.07239}}].

\bibitem{Bak:2018txn}
D.~Bak, C.~Kim and S.-H. Yi, \emph{{Bulk view of teleportation and traversable
  wormholes}}, \href{https://doi.org/10.1007/JHEP08(2018)140}{\emph{JHEP}
  {\bfseries 08} (2018) 140}
  [\href{https://arxiv.org/abs/1805.12349}{{\ttfamily 1805.12349}}].

\bibitem{Ahn:2022suv}
B.~Ahn, S.-E. Bak, V.~Jahnke and K.-Y. Kim, \emph{{Traversable wormholes via a
  double trace deformation involving U(1) conserved current operators}},
  \href{https://doi.org/10.1103/PhysRevD.109.066016}{\emph{Phys. Rev. D}
  {\bfseries 109} (2024) 066016}
  [\href{https://arxiv.org/abs/2206.03434}{{\ttfamily 2206.03434}}].

\bibitem{Khairunnisa:2025ljk}
F.~Khairunnisa, H.~L. Prihadi, M.~Z. Djogama, D.~Dwiputra and F.~P. Zen,
  \emph{{Traversable wormhole with double trace deformations via gravitational
  shear and sound channels}},
  \href{https://arxiv.org/abs/2511.09815}{{\ttfamily 2511.09815}}.

\bibitem{Djogama:2026pmd}
M.~Z. Djogama, F.~Khairunnisa, H.~L. Prihadi and F.~P. Zen, \emph{{Kerr/CFT
  Traversable Wormhole with Fermionic Double-Trace Deformation}},
  \href{https://arxiv.org/abs/2605.05011}{{\ttfamily 2605.05011}}.

\bibitem{Freedman:2016zud}
M.~Freedman and M.~Headrick, \emph{{Bit threads and holographic entanglement}},
  \href{https://doi.org/10.1007/s00220-016-2796-3}{\emph{Commun. Math. Phys.}
  {\bfseries 352} (2017) 407}
  [\href{https://arxiv.org/abs/1604.00354}{{\ttfamily 1604.00354}}].

\bibitem{Takayanagi:2017knl}
T.~Takayanagi and K.~Umemoto, \emph{{Entanglement of purification through
  holographic duality}},
  \href{https://doi.org/10.1038/s41567-018-0075-2}{\emph{Nature Phys.}
  {\bfseries 14} (2018) 573}
  [\href{https://arxiv.org/abs/1708.09393}{{\ttfamily 1708.09393}}].

\bibitem{Czech:2012bh}
B.~Czech, J.~L. Karczmarek, F.~Nogueira and M.~Van~Raamsdonk, \emph{{The
  Gravity Dual of a Density Matrix}},
  \href{https://doi.org/10.1088/0264-9381/29/15/155009}{\emph{Class. Quant.
  Grav.} {\bfseries 29} (2012) 155009}
  [\href{https://arxiv.org/abs/1204.1330}{{\ttfamily 1204.1330}}].

\bibitem{Wall:2012uf}
A.~C. Wall, \emph{{Maximin Surfaces, and the Strong Subadditivity of the
  Covariant Holographic Entanglement Entropy}},
  \href{https://doi.org/10.1088/0264-9381/31/22/225007}{\emph{Class. Quant.
  Grav.} {\bfseries 31} (2014) 225007}
  [\href{https://arxiv.org/abs/1211.3494}{{\ttfamily 1211.3494}}].

\bibitem{Headrick:2014cta}
M.~Headrick, V.~E. Hubeny, A.~Lawrence and M.~Rangamani, \emph{{Causality {\&}
  holographic entanglement entropy}},
  \href{https://doi.org/10.1007/JHEP12(2014)162}{\emph{JHEP} {\bfseries 12}
  (2014) 162} [\href{https://arxiv.org/abs/1408.6300}{{\ttfamily 1408.6300}}].

\bibitem{Betzios:2023obs}
P.~Betzios and O.~Papadoulaki, \emph{{Wilson loops and wormholes}},
  \href{https://doi.org/10.1007/JHEP03(2024)066}{\emph{JHEP} {\bfseries 03}
  (2024) 066} [\href{https://arxiv.org/abs/2311.09289}{{\ttfamily
  2311.09289}}].

\bibitem{Betzios:2019rds}
P.~Betzios, E.~Kiritsis and O.~Papadoulaki, \emph{{Euclidean Wormholes and
  Holography}}, \href{https://doi.org/10.1007/JHEP06(2019)042}{\emph{JHEP}
  {\bfseries 06} (2019) 042}
  [\href{https://arxiv.org/abs/1903.05658}{{\ttfamily 1903.05658}}].

\bibitem{Betzios:2021fnm}
P.~Betzios, E.~Kiritsis and O.~Papadoulaki, \emph{{Interacting systems and
  wormholes}}, \href{https://doi.org/10.1007/JHEP02(2022)126}{\emph{JHEP}
  {\bfseries 02} (2022) 126}
  [\href{https://arxiv.org/abs/2110.14655}{{\ttfamily 2110.14655}}].

\bibitem{Low:2001bw}
I.~Low and A.~V. Manohar, \emph{{Spontaneously broken space-time symmetries and
  Goldstone's theorem}},
  \href{https://doi.org/10.1103/PhysRevLett.88.101602}{\emph{Phys. Rev. Lett.}
  {\bfseries 88} (2002) 101602}
  [\href{https://arxiv.org/abs/hep-th/0110285}{{\ttfamily hep-th/0110285}}].

\bibitem{Etheredge:2026rio}
M.~Etheredge, M.~Reece, T.~Rudelius and C.~Tudball, \emph{{Sharpening the
  Supersymmetric Axion Weak Gravity Conjecture}},
  \href{https://arxiv.org/abs/2605.22912}{{\ttfamily 2605.22912}}.

\bibitem{Harlow:2015lma}
D.~Harlow, \emph{{Wormholes, Emergent Gauge Fields, and the Weak Gravity
  Conjecture}}, \href{https://doi.org/10.1007/JHEP01(2016)122}{\emph{JHEP}
  {\bfseries 01} (2016) 122}
  [\href{https://arxiv.org/abs/1510.07911}{{\ttfamily 1510.07911}}].

\bibitem{Guica:2015zpf}
M.~Guica and D.~L. Jafferis, \emph{{On the construction of charged operators
  inside an eternal black hole}},
  \href{https://doi.org/10.21468/SciPostPhys.3.2.016}{\emph{SciPost Phys.}
  {\bfseries 3} (2017) 016} [\href{https://arxiv.org/abs/1511.05627}{{\ttfamily
  1511.05627}}].

\bibitem{Geng:2025byh}
H.~Geng, \emph{{Making the Case for Massive Islands}},
  \href{https://arxiv.org/abs/2509.22775}{{\ttfamily 2509.22775}}.

\bibitem{Geng:2025gqu}
H.~Geng, D.~Jafferis, P.~Shrivastava and N.~Tata, \emph{{The Fate of
  Information Localizability and Holography in Quantum Gravity}},
  \href{https://arxiv.org/abs/2512.18912}{{\ttfamily 2512.18912}}.

\bibitem{Geng:2025gns}
H.~Geng, J.~Huertas, A.~Karch, L.~Randall and D.~Thomas, \emph{{Wet Hair:
  Global Symmetries in Entanglement Islands}},
  \href{https://arxiv.org/abs/2512.11025}{{\ttfamily 2512.11025}}.

\bibitem{Geng:2025rov}
H.~Geng, \emph{{The mechanism behind the information encoding for islands}},
  \href{https://doi.org/10.1007/JHEP03(2026)037}{\emph{JHEP} {\bfseries 03}
  (2026) 037} [\href{https://arxiv.org/abs/2502.08703}{{\ttfamily
  2502.08703}}].

\bibitem{Kiritsis:2019wyk}
E.~Kiritsis and A.~Tsouros, \emph{{De Sitter versus Anti de Sitter flows and
  the (super)gravity landscape}},
  \href{https://doi.org/10.1007/JHEP08(2023)126}{\emph{JHEP} {\bfseries 08}
  (2023) 126} [\href{https://arxiv.org/abs/1901.04546}{{\ttfamily
  1901.04546}}].

\bibitem{Kiritsis:2025ytb}
E.~Kiritsis, S.~Morales-Tejera and C.~Rosen, \emph{{de Sitter versus Anti de
  Sitter flows and the (super)gravity landscape: Part II}},
  \href{https://arxiv.org/abs/2510.12373}{{\ttfamily 2510.12373}}.

\bibitem{Khavaev:2000gb}
A.~Khavaev and N.~P. Warner, \emph{{A Class of N=1 supersymmetric RG flows from
  five-dimensional N=8 supergravity}},
  \href{https://doi.org/10.1016/S0370-2693(00)01228-4}{\emph{Phys. Lett. B}
  {\bfseries 495} (2000) 215}
  [\href{https://arxiv.org/abs/hep-th/0009159}{{\ttfamily hep-th/0009159}}].

\bibitem{code}
{\emph{https://github.com/JosemaGIT-0702/Wormhole-Spectrum/tree/main} }.

\end{thebibliography}\endgroup

\end{document}